\documentclass[manuscript, acmsmall,screendui]{acmart}
\usepackage[normalem]{ulem}
\useunder{\uline}{\ul}{}

\usepackage{multirow} 

\setcopyright{acmlicensed}
\copyrightyear{2018}
\acmYear{2018}
\acmDOI{XXXXXXX.XXXXXXX}

\acmConference[Conference acronym 'XX]{Make sure to enter the correct
  conference title from your rights confirmation emai}{June 03--05,
  2018}{Woodstock, NY}
\acmISBN{978-1-4503-XXXX-X/18/06}

\begin{document}


\title[A Systematic Review of Presence in Virtual Reality Across Tasks and Disciplines]{Towards Enhanced Learning through Presence: A Systematic Review of Presence in Virtual Reality Across Tasks and Disciplines}
\author{Zheng Wei}
\email{zwei302@connect.ust.hk}
\orcid{0000-0001-7444-2547}
\affiliation{%
  \institution{The Hong Kong University of Science and Technology}
  \city{Hong Kong}
  \state{Hong Kong SAR}
  \country{China}
}

\author{Junxiang Liao}
\affiliation{%
  \institution{The Hong Kong University of Science and Technology (Guangzhou)}
  \city{Guangzhou}
  \country{China}}
\email{jliao659@connect.hkust-gz.edu.cn}

\author{Lik-Hang Lee}
\affiliation{%
  \institution{The Hong Kong Polytechnic University}
  \city{Hong Kong}
  \country{China}}
\email{lik-hang.lee@polyu.edu.hk}

\author{Huamin Qu}
\authornote{Qu is the corresponding author.}
\affiliation{%
  \institution{The Hong Kong University of Science and Technology}
  \city{Hong Kong}
  \country{China}}
\email{huamin@cse.ust.hk}

\author{Xian Xu}
\authornote{Xian is the corresponding author.}
\affiliation{%
  \institution{The Hong Kong University of Science and Technology}
  \city{Hong Kong}
  \country{China}}
\email{xianxu@ust.hk}

\renewcommand{\shortauthors}{Wei et al.}
\begin{abstract}
The rising interest in Virtual Reality (VR) technology has sparked a desire to create immersive learning platforms capable of handling various tasks across environments. Through immersive interfaces, users can engage deeply with virtual environments, enhancing both learning outcomes and task performance. In fields such as education, engineering, and collaboration, presence has emerged as a critical factor influencing user engagement, motivation, and skill mastery. This review provides a comprehensive examination of the role of presence across different tasks and disciplines, exploring how its design impacts learning outcomes. Using a systematic search strategy based on the \textit{PRISMA} method, we screened 2,793 articles and included 78 studies that met our inclusion criteria. We conducted a detailed classification and analysis of different types of presence in VR environments, including spatial presence, social presence, co-presence, self-presence, and cognitive presence. This review emphasizes how these varied types of presence affect learning outcomes across tasks and fields, and examines how design elements and interaction techniques shape presence and subsequently impact learning outcomes. We also summarize trends and future directions, identifying research gaps and opportunities to improve learning outcomes by enhancing presence in VR environments, thus offering guidance and insight for future research on VR presence and learning effectiveness.
\end{abstract}
\ccsdesc[500]{Human-centered computing~Virtual reality}
\keywords{Immersive Learning, Virtual Reality Education, Presence Theory, Human-Computer Interaction}
\begin{teaserfigure}
  \includegraphics[width=0.9\textwidth]{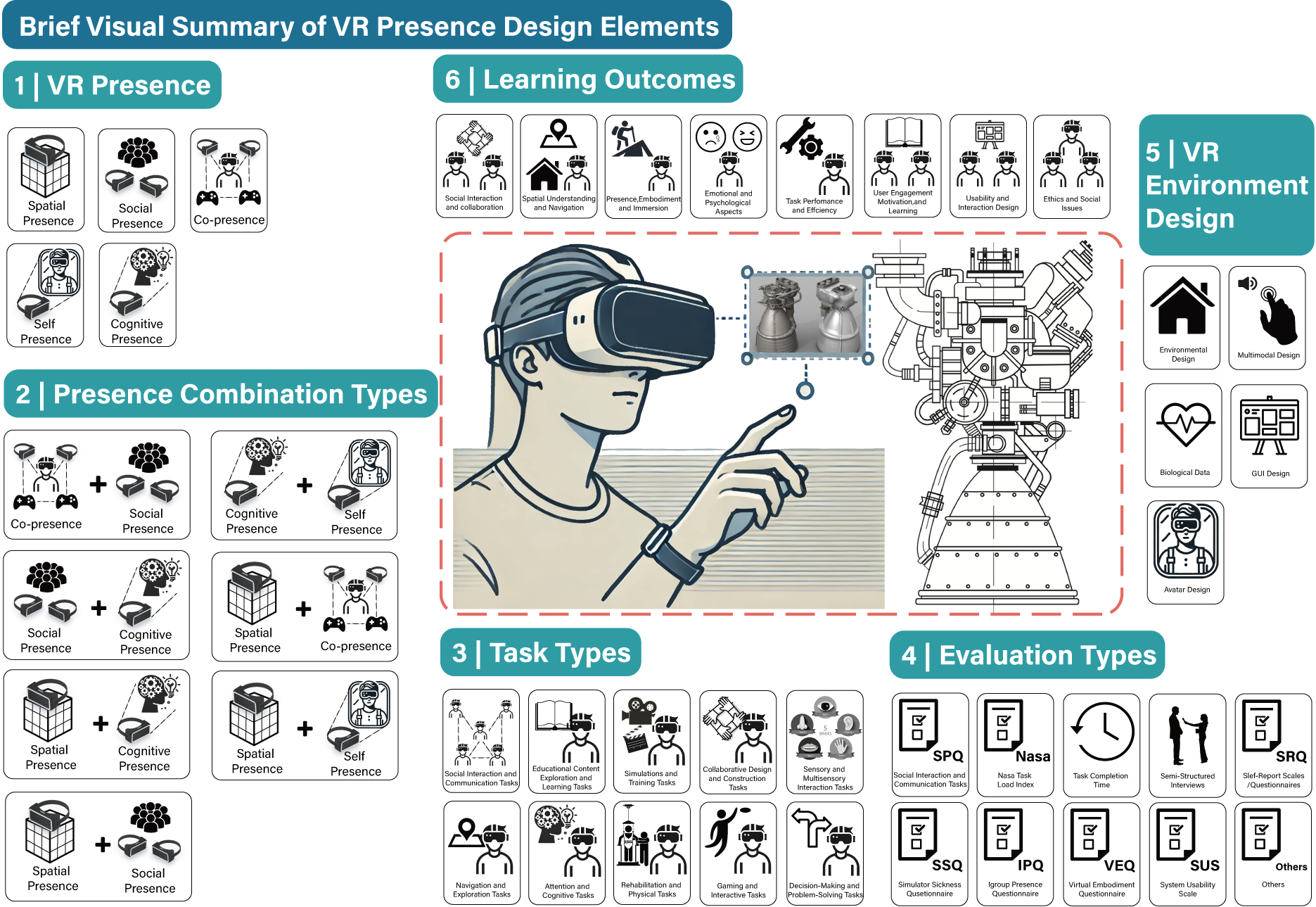}
  \caption{Brief visual summary of key VR presence design elements, highlighting six main categories: VR Presence, Presence Combination Types, Task Types, Evaluation Types, VR Environment Design, and Learning Outcomes.}
  \label{fig:teaser}
\end{teaserfigure}
\maketitle

\section{Introduction}

\subsection{Overview of Presence in Virtual Reality Learning Environments}

\subsubsection{Virtual Environment}

The `Virtual Environment' refers to the immersive digital space where learning activities occur. It is often designed using VR technologies to simulate real-world settings or create entirely new environments that facilitate educational experiences (see Figure \ref{title}). This environment is crucial for inducing various types of presence, such as spatial presence and self-presence. For example, in a virtual chemistry lab, learners can perform experiments that might be dangerous or impractical in a physical lab. The virtual environment provides interactive elements, 3D models, and simulations that allow learners to engage deeply with the content. The design of this space should consider factors like realism, interactivity, and sensory feedback to effectively enhance the learner's sense of 'being there'.

\subsubsection{Learner Experience}
The `Learner Experience' encompasses the cognitive and emotional responses of the user as they interact within the virtual environment (see Figure \ref{title}). This includes their sense of presence—spatial, social, co-presence, self-presence, and cognitive presence—as well as engagement, motivation, and learning outcomes. For example, when learners collaborate with virtual avatars or other real learners within the VR environment, their social presence and co-presence are heightened, potentially leading to improved learning outcomes. The learner experience is influenced by the design of the virtual environment, the interaction modalities available, and how well the virtual tasks align with educational objectives.

\subsection{Objective of the Survey}

To provide a comprehensive understanding of the factors influencing presence and learning outcomes in VR environments, we have conducted a detailed synthesis and analysis of the relevant literature. The objective of this survey is to categorize and examine these influencing factors along several key dimensions.

First, we analyzed the factors influencing presence and learning outcomes in virtual reality environments. Based on our systematic review, we have categorized the factors influencing presence and learning outcomes in VR environments along the following dimensions:

\subsubsection{Design Elements of the Virtual Environment}

The architectural and aesthetic aspects of the virtual environment play a crucial role in fostering various types of presence. High-fidelity scene rendering, dynamic scene interactions, and adaptive environmental feedback are essential for creating immersive and realistic settings. These design elements not only enhance spatial presence but also support cognitive and social engagements by providing a coherent and responsive virtual space for learners to interact with.

\subsubsection{Types of Presence and Their Interactions}

Understanding the interplay between different types of presence—spatial, social, co-presence, self-presence, and cognitive presence—is essential for designing effective VR learning environments. Our analysis reveals that combinations of these presence types can synergistically enhance various aspects of the learning experience. For instance, the intersection of spatial and social presence is particularly effective in collaborative learning tasks, while spatial and cognitive presence together support simulations and training activities that require deep cognitive engagement.

\subsubsection{Multimodal Sensory Feedback}

Effective integration of multiple sensory modalities—visual, auditory, haptic, and olfactory—significantly enhances the sense of immersion and presence. Visual cues such as high-resolution textures and real-time object feedback improve spatial awareness, while spatial audio and haptic feedback facilitate realistic interactions and emotional resonance. The synergistic effect of these sensory inputs contributes to a more engaging and effective learning experience by catering to different sensory preferences and reducing cognitive load.

\subsubsection{Avatar and Identity Representation}

The design and customization of avatars are pivotal in enhancing social presence and self-presence. High-fidelity anthropomorphic avatars that accurately reflect users' facial expressions and body movements enable more natural and meaningful social interactions. Additionally, the ability to customize avatars for self-expression fosters a stronger sense of ownership and personal identification within the virtual environment, which can lead to increased motivation and engagement in learning tasks.

\subsubsection{Interaction Modalities and User Interfaces}

The design of graphical user interfaces (GUI) and interaction mechanisms directly impacts how users engage with the virtual environment. Intuitive and responsive interfaces, real-time information feedback, and interactive control systems facilitate smoother and more efficient interactions. These elements are critical for minimizing cognitive strain and enabling learners to focus on task-related activities, thereby improving task performance and learning outcomes.

\subsubsection{Biometric Data Integration}

Incorporating biometric data such as heart rate, eye-tracking, and EEG provides objective measures of user states, offering deeper insights into their emotional and cognitive experiences. This integration allows for adaptive learning environments that can respond to users' physiological states in real-time, thereby optimizing the balance between immersion and cognitive load. Biometric feedback can also enhance the reliability of evaluations by correlating physiological responses with self-reported measures of presence and engagement.

\subsubsection{Task Types and Educational Objectives}

The nature of the tasks performed within the VR environment significantly influences the required presence types and the corresponding learning outcomes. Tasks categorized as social interaction and communication, collaborative design and construction, gaming and interactive activities, and educational content exploration each demand different configurations of presence and interaction strategies. Aligning the design of the virtual environment and presence-enhancing factors with the specific educational objectives and task requirements ensures that the VR experience effectively supports learning and skill acquisition.

\subsubsection{Evaluation Methods and Measurement Metrics}

The methodologies employed to assess presence and learning outcomes play a critical role in understanding their relationship. Quantitative measures such as task completion time, accuracy, standardized questionnaires (e.g., IPQ, NASA-TLX), and physiological data provide objective assessments of user experiences and performance. Mixed methods approaches that combine quantitative and qualitative data offer comprehensive insights into how presence influences learning processes and outcomes, enabling more nuanced interpretations and targeted improvements in VR educational designs.

\subsection{Existing Surveys}
Several surveys and literature reviews have been conducted to explore various aspects of Virtual Reality (VR), including its technological advancements, applications in remote collaboration, and implications in Human-Computer Interaction (HCI). Skarbez et al. \cite{skarbez2017survey} provided an extensive review of presence and related constructs, focusing on definitions, theoretical models, and measurement techniques. Their work primarily concentrated on the conceptual underpinnings of presence and the factors that influence it within VR environments.

Similarly, Cummings and Bailenson \cite{cummings2016immersive} conducted a meta-analysis examining how immersive VR technology affects user presence. Their study evaluated the relationship between levels of immersion and the sense of presence, providing valuable insights into how technological components contribute to the user's immersive experience. However, both of these surveys predominantly addressed the technological and psychological aspects of presence without delving deeply into how different types of presence specifically impact learning outcomes across various tasks and disciplines.


In the fields of HCI and education, most researchers have focused on the application of VR technologies to enhance user experience and learning outcomes (e.g., Kaminska et al., \cite{kaminska2019virtual}; Makinen et al., \cite{makinen2022user}; Kaplan et al., \cite{kaplan2021effects}). These studies often emphasize the effectiveness of VR in simulating real-world scenarios and the technological aspects of VR applications. However, they tend to overlook how different types of presence can influence learning outcomes and user engagement in educational settings.

Although these existing surveys have made significant contributions to understanding presence in VR, there remains a gap in the literature regarding how to comprehensively analyze how various types of presence, such as spatial presence, social presence, co-presence, self-presence, and cognitive presence—affect learning outcomes in educational and collaborative tasks. Our research aims to fill this gap by systematically examining the design elements and interaction techniques that influence these types of presence, and how they impact learning effectiveness across different fields. By undertaking this approach, we offer a more comprehensive understanding that integrates the theoretical concepts of presence with their practical applications in virtual reality learning environments.


\subsection{Structure of the Survey}

The remainder of this article is structured as follows: Sec. \ref{section2} thoroughly explains the Methodology employed for this survey, detailing the search strategy, inclusion and exclusion criteria, and data extraction and analysis methods. Sec. \ref{section3} presents the Results and Descriptive Statistics, offering an overview of the included articles and focusing on the different types of presence and associated tasks. In Sec. \ref{section4}, we provide an in-depth Discussion concerning the development and recent advancements in VR-based training and collaboration. This includes the moderating effect of design factors on presence and learning outcomes, with subtopics on multi-sensory feedback design, virtual character and emotional expression design, self-presence and body-perception design, and feedback mechanisms and interaction design. We also explore the multidimensional influence of presence in engineering and collaborative tasks, discussing its direct and indirect effects, dynamic adjustment and adaptation, and its universality and specificity in interdisciplinary applications. Practical application and design suggestions are also provided. Sec. \ref{section5} discusses the challenges and future directions of the research, covering technical limitations, scalability, ethical and social concerns, the potential of VR in distance learning, and its practical deployment. Sec. \ref{section6} concludes the article, summarizing the main findings and contributions. 

\section{METHODOLOGY}\label{section2}
\begin{figure}[ht]
    \centering
    \includegraphics[width=13cm]{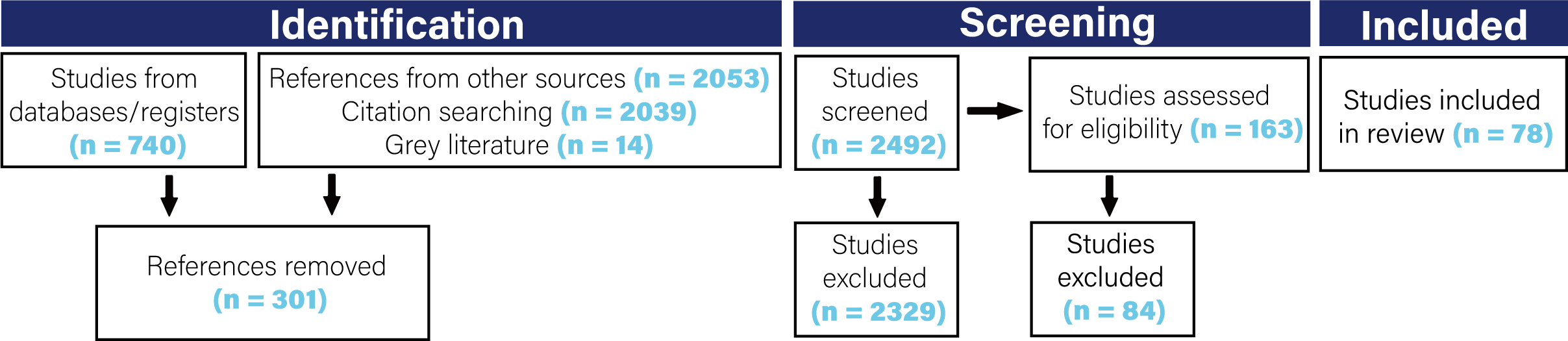}
    \caption{Systematic review process using \textit{PRISMA}}
    \label{PRISMA}
\end{figure}

To ensure methodological rigor and enhance transparency in the literature review process, we implemented the \textit{PRISMA} (Preferred Reporting Items for Systematic Reviews and Meta-Analyses) framework, as advocated by Sarkis-Onofre et al. \cite{sarkis2021properly}. The detailed \textit{PRISMA} flowchart is presented in Figure \ref{PRISMA}. Multiple authors conducted the review collaboratively using an online tool named Covidence\footnote{\url{https://app.covidence.org/reviews/active}}. Upon completing the initial search, we commenced a structured screening of 2793 studies. After removing 301 duplicate records, 2492 studies remained for further consideration. Screening by title and abstract led to the exclusion of 2329 studies that did not satisfy our criteria. Following a thorough full-text evaluation, an additional 84 studies were excluded. Specific inclusion and exclusion criteria are elaborated in Sec. \ref{section2.2}. The literature review was conducted in two stages to capture recent developments in the field. The first round of data extraction occurred in August 2024, with a second round in December 2024. This phased approach enabled us to incorporate the latest research and ensure a thorough examination of the literature up to December 2024. Ultimately, 78 studies were retained for data extraction and in-depth analysis within the present study.

\subsection{Search Strategy}
To conduct a comprehensive and systematic review of the literature on presence in VR learning environments, we developed a rigorous search strategy following the \textit{PRISMA} guidelines. 

\subsubsection{Keywords}
The selection of keywords was crucial to capture the breadth of research related to presence in VR and its impact on learning outcomes across different tasks and disciplines. We identified primary keywords based on the theoretical frameworks of presence and VR in educational contexts, as well as commonly used terms in the literature. The main keywords and phrases used in our search included: ``Presence'', ``Sense of Presence'', ``Spatial Presence'', ``Social Presence'', ``Co-presence'', ``Self-presence'', ``Cognitive Presence'', ``Virtual Reality'', ``Immersive Learning'', ``User Experience'', ``Presence Enhancement Strategies'' and the abbreviations ``VR'', ``HCI'', ``VR Training Systems'', ``Embodiment in VR'', ``VR Learning Outcome'' and  ``VR Educational Technology'', ``VR Collaborative Learning'', ``VR Interaction Design'', ``VR Skill Acquisition'' and ``VR Cognitive Load''.

We combined these keywords using \textit{Boolean operators} \cite{dinet2004searching} to refine the search and ensure the inclusion of relevant studies. For example: ((``Presence'' OR ``Sense of Presence'') AND (``User Experience'' OR ``Immersive Learning'')) AND (``VR Training Systems'' OR ``VR Educational Technology'' OR ``VR Collaborative Learning''). Wildcards and truncation symbols were also used where appropriate to capture variations of keywords (e.g., ``educat*'' to include ``education,'' ``educational,'' ``educate'').

\subsubsection{Databases}
Building upon an initial research list established through keyword searches, we further employed a backward citation method to conduct an in-depth exploration of relevant literature. Specifically, we selected pertinent research reports from the proceedings of premier conferences such as \textit{IEEE VR}, \textit{IEEE ISMAR}, \textit{ACM MM}, \textit{ACM UIST}, \textit{ACM VRST}, \textit{ACM CSCW}, and \textit{ACM CHI}. These conferences were chosen due to their reputation as leading forums for advancements in virtual reality, augmented reality, multimedia, user interface design, and human-computer interaction, ensuring access to high-quality and cutting-edge research. Additionally, we reviewed a variety of high-impact journal articles covering fields including human-computer interaction, human factors engineering design, and educational sciences, such as \textit{IEEE TVCG} and \textit{TOCHI}. These journals were selected for their rigorous peer-review processes and their prominence in disseminating influential studies within their respective disciplines. To complement this approach, we conducted a snowball search on the \textit{Google Scholar} platform by examining the reference lists of related articles to identify additional relevant publications. Furthermore, we utilized the online tool \textit{Connected Papers} \footnote{\url{https://www.connectedpapers.com/}} to identify and expand the scope of related papers until no new relevant literature was found. This comprehensive literature search methodology ensured thorough coverage of pertinent research, effectively identifying the primary literature resources closely related to our research questions.

\subsection{Inclusion and Exclusion Criteria} \label{section2.2}

During the screening and eligibility phases of the \textit{PRISMA} review, we meticulously developed appropriate inclusion and exclusion criteria to guide the selection process. Our analysis focused on articles that met one or more inclusion criteria while ensuring the exclusion of any studies that met at least one exclusion criterion, thereby guaranteeing a rigorous and comprehensive review of the relevant literature. In this process, we identified multiple studies related to user presence. Specifically, the studies included in the analysis must satisfy the following five criteria:

1) Include the manipulation of one or more specific types of presence (e.g., spatial presence, social presence, co-presence, self-presence, and cognitive presence); 

2) Contain at least one self-reported measure of user presence;

3) Provide sufficient details to determine the relative immediacy of the compared conditions;


4) Be published within the last five years, to ensure the timeliness and relevance of the research;

5) Research on how design elements or interaction technologies that affect types of presence influence learning outcomes in educational or collaborative tasks;

By adhering to these guidelines, we ensure that the included studies are highly relevant and timely, thereby providing a solid foundation for our review.

\subsection{Data Extraction and Analysis}
To extract relevant information from the included articles, we developed a data extraction protocol to systematically capture key aspects of various types of presence in educational and collaborative tasks. Initially, the first author of this review pseudo-randomly selected 10 articles and established preliminary data items based on the relevant elements identified within them. Subsequently, all authors evaluated the initial data extraction criteria and refined them into the final version presented in Table \ref{Table1}. The first and second authors independently extracted data from each article according to the finalized criteria. In cases of data discrepancies, the authors reached a consensus through discussion. 

The data extraction items (\textbf{DE1-DE12}) cover various aspects of the studies analyzed. General information (\textbf{DE1-DE3}) includes the study identifier (author, source, and year), title, and keywords. The type of presence (\textbf{DE4}) encompasses spatial presence, social presence, co-presence, self-presence, and cognitive presence. Task types (\textbf{DE5}) involve 10 categories, such as collaborative design and construction tasks, while presence combinations (\textbf{DE6}) include eight variations, such as co-presence \& social presence. Evaluation methods (\textbf{DE7}) consist of 10 types, including task completion time and simulator sickness questionnaire (SSQ). The VR environment design (\textbf{DE8}) covers six types, such as avatar design and environment design, and collaboration type (\textbf{DE9}) distinguishes between multiplayer collaboration and single-user settings. Learning outcomes (\textbf{DE10}) focus on eight indicators, such as social interaction, emotional and psychological aspects, and collaboration. The number of participants (\textbf{DE11}) ensures representativeness and statistical power, while the domain (\textbf{DE12}) identifies whether the study belongs to social sciences or engineering fields.

\begin{table}[ht]
\tiny
\centering
\caption{Data Extraction Rubric for the Selected 78 Articles}
\scalebox{0.85}{
\begin{tabular}{cll}
\hline
ID           & \multicolumn{1}{c}{\textbf{Data Extraction}} & \multicolumn{1}{c}{\textbf{Description}}                                                                                                                                        \\ \hline
DE1          & Study identifier                             & Includes study identifier (author, source and year)                                                                                                                             \\
DE2          & Title                                        & Title of the paper                                                                                                                                                              \\
DE3          & Keywords                                     & Keywords of the paper                                                                                                                                                           \\
DE4          & Type of Presence                             & Encompasses spatial presence, social presence, co-presence, self-presence, and cognitive presence.                                                                              \\
DE5          & Types of  Tasks                              & There are 10 types of tasks, e.g. collaborative design and construction tasks, etc.                                                                                             \\
DE6 & Presence Combination                         & The presence combination contains 8 types, e.g. co-presence \& social presence, etc.                                                                                             \\
DE7 & Evaluation Types                             & There are 10 types, e.g. task completion time, simulator sickness questionnaire (SSQ), etc.                                                                                     \\
DE8          & VR Environment Design                        & There are six types, e.g. avatar design, environment design, etc.                                                                                                                 \\
DE9          & Collaboration Type                           & Includes multiplayer collaboration and single user.                                                                                                                             \\
DE10         & Learning Outcomes                            & There are eight types, e.g. social interaction and collaboration, emotional and psychological aspects, etc.                                                                         \\
DE11         & Participant Numbers                          & \begin{tabular}[c]{@{}l@{}}Refers to the number of participants in the study to evaluate the representativeness and statistical \\ power of the research findings.\end{tabular} \\
DE12         & Domain                                       & Indicates whether the study belongs to the social sciences or engineering fields.                                                                                               \\ \hline
\end{tabular}
}
\label{Table1}
\end{table}

\section{RESULTS AND DESCRIPTIVE STATISTICS}\label{section3}

\subsection{Overview of Included Articles}
\begin{figure}[ht]
    \centering
    \includegraphics[width=13cm]{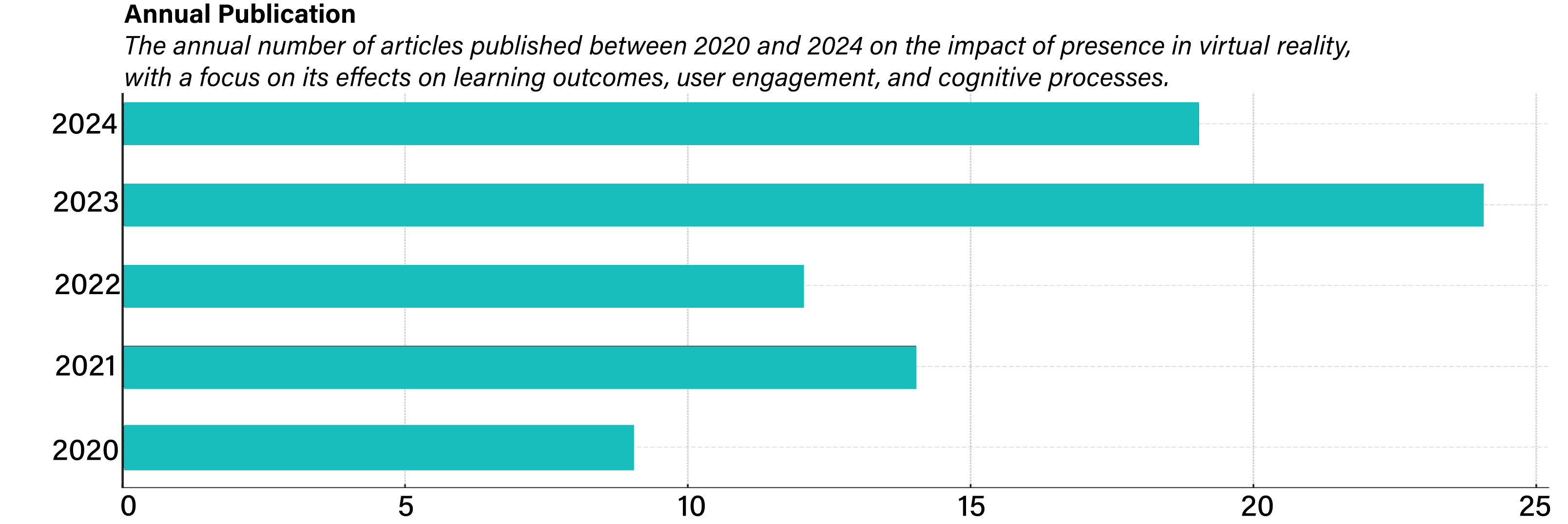}
    \caption{Articles per year (N=78)}
    \label{annual}
\end{figure}

We meticulously extracted relevant information from the selected 78 articles. The chosen publications span from 2020 to 2024, with the annual publication counts illustrated in Figure \ref{annual}. These articles originate from renowned conferences and journals in the fields of computer graphics and human-computer interaction (HCI), including the IEEE Conference on Virtual Reality and 3D User Interfaces (VR), ACM Conference on Human Factors in Computing Systems (CHI), ACM Multimedia (MM), ACM Symposium on User Interface Software and Technology (UIST), IEEE International Symposium on Mixed and Augmented Reality (ISMAR), ACM Symposium on Virtual Reality Software and Technology (VRST), ACM Computer-Supported Cooperative Work \& Social Computing (CSCW), IEEE Transactions on Visualization and Computer Graphics (TVCG), ACM Transactions on Computer-Human Interaction (TOCHI), and the International Journal of Human-Computer Studies (IJHCS).

Subsequently, we conducted statistical analyses of these data and presented them through charts, deriving additional qualitative insights through iterative analysis. Detailed data extraction for all literature can be found in Appendix \ref{A}.

Over the past five years, the number of publications related to this research topic has generally increased, indicating a significant rise in attention and an enhancement in the field's importance. This upward trend is primarily attributed to the continuous development of virtual reality technology and the popularity of concepts such as the ``Metaverse''. Notably, the field reached its peak in publication output in 2023, reflecting a high level of research activity during that year.

\begin{figure}[ht]
    \centering
    \includegraphics[width=8cm]{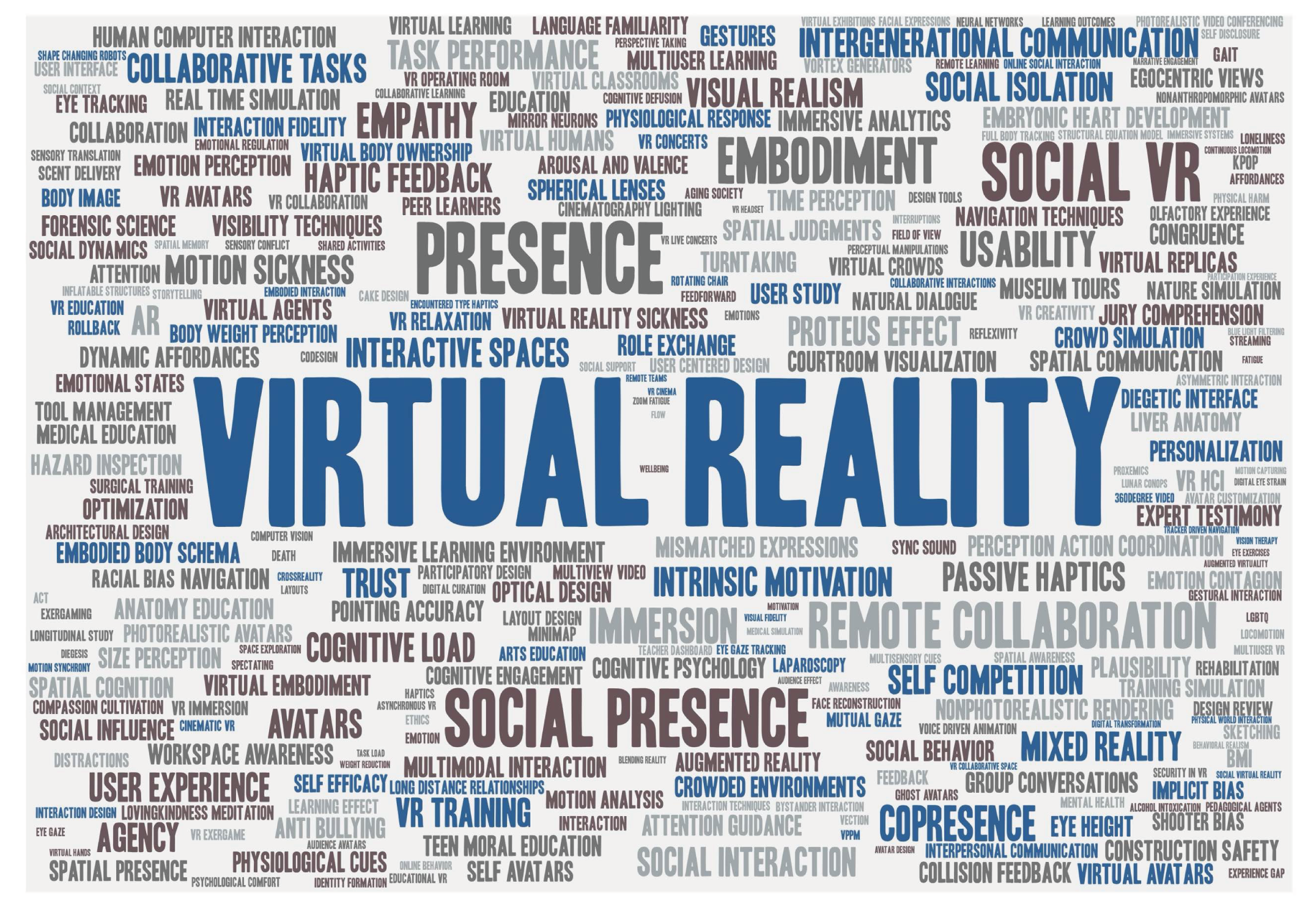}
    \caption{Frequency distribution of keywords in included articles.}
    \label{key}
\end{figure}

These articles primarily focus on the fields of virtual reality and remote interaction. Our research stands out by systematically investigating various types of ``presence'' in virtual reality (e.g., spatial presence, social presence, co-presence, self-presence, and cognitive presence), particularly how they influence remote collaboration and educational and collaborative tasks within virtual environments through design elements and interaction technologies. The word cloud illustrates that ``virtual reality'' is the most frequently occurring keyword, reflecting the current research area's strong emphasis on virtual reality technology. Additionally, keywords such as ``remote collaboration,'' ``social presence,'' and ``immersion'' also appear frequently, indicating that the research focus is on collaboration technologies, user experience, and interaction quality within virtual environments. Keywords related to ``avatars'' highlight the importance of identity expression and personalization for users in virtual settings, while keywords like ``emotion'' and ``trust'' emphasize the complexity of cultural and social interactions in virtual environments. Figure \ref{key} visually presents the frequency distribution of keywords in the form of a word cloud, providing a visual reference for the hotspots and trends in this research field.

\subsection{Presence types and Tasks}

In our analysis of the reviewed studies, we found a close and complex relationship between types of presence and task types. Analyzing these data provides a comprehensive understanding of the application trends of VR technology in education and collaborative tasks.

\begin{figure}[ht]
    \centering
    \includegraphics[width=9cm]{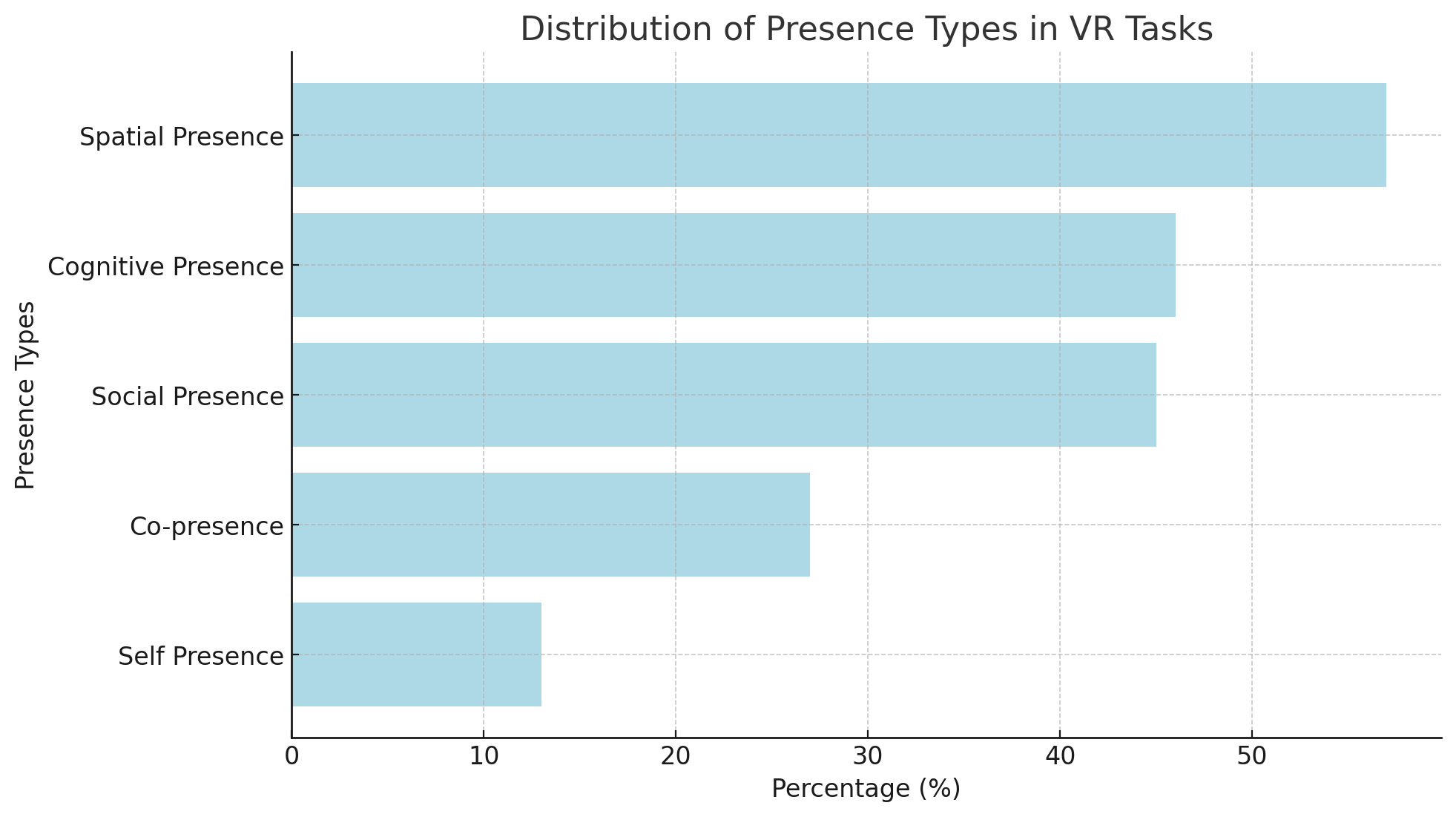}
    \caption{Statistics on the types of presence used in the studies (some studies involve multiple types of presence simultaneously.}
    \label{presence type}
\end{figure}

\subsubsection{Presence Combination Types}
In analyzing the relevant literature, we found that most articles reference more than one dimension of presence - typically presenting one primary dimension (e.g., spatial presence) while also discussing other supporting dimensions. To ensure completeness and clarity, every presence dimension identified in these studies is reported and analyzed in the following subsections.

Spatial presence was observed in 57\% (45/78) of the studies, highlighting its importance in enabling users to perceive their position and spatial relationships within a virtual environment \cite{bhargava2023empirically}\cite{wijayanto2023comparing}\cite{schott2021vr}\cite{peck2021evidence}\cite{wagner2021effect}. This type of presence is particularly critical for navigation and exploration tasks, simulation and training tasks, and rehabilitation and bodily tasks. For example, in the study by Bhargava et al. \cite{bhargava2023empirically}, participants performed a road crossing task in a virtual environment, requiring accurate spatial perception to evaluate obstacle dimensions and distances. Enhanced spatial presence allowed participants to navigate more naturally, improving task success rates. Similarly, in Wei et al.'s research \cite{wei2023feeling}, participants learned professional lighting techniques in a virtual film studio, accurately perceiving the effects of simulated lighting. The strengthened spatial presence allowed participants to develop a more concrete understanding of abstract lighting effects \cite{xu2023cinematography,xu2024transforming}. 

Cognitive presence was observed in 46\% (36/78) of the studies. Cognitive presence focuses on users' deep thinking and knowledge construction by users within the virtual environment, which is essential for decision-making and problem-solving tasks, attention and cognition tasks, and educational content exploration and learning tasks \cite{shigyo2024vr}\cite{bueno2021effects}\cite{bellgardt2023virtual}\cite{schott2023cardiogenesis4d}. For example, Bellgardt et al.'s virtual optical platform \cite{bellgardt2023virtual} enabled students to adjust spherical lens layouts in real-time, fostering a deeper understanding of optical principles. 

The social presence appeared in 45\% (35/78) of the studies, emphasizing its critical role in the success of collaborative tasks. Tasks such as collaborative design and construction, social interaction and communication, and exploration of educational content are heavily based on social presence \cite{wong2023comparing}\cite{lee2024may}\cite{espositi2024room}\cite{wang2023designing}\cite{mei2021cakevr}. The research of Wong et al. \cite{wong2023comparing} demonstrated that improved social presence contributed to successful collaboration during tasks.

Copresence, observed in 27\% (21/78) of the studies, improved synchronous interaction and emotional resonance among users, particularly in games and interactive tasks and social interaction and communication tasks \cite{tian2023using}\cite{volonte2021effects}\cite{wei2023bridging}\cite{abramczuk2023meet} .In Tian et al.'s study \cite{tian2023using}, virtual replicas were utilized to improve remote collaboration in mixed reality environments. Participants worked together to make objects, with increased social presence that improved communication efficiency and collaboration satisfaction. 

Self-presence had lower frequencies, appearing in 13\% (10/78)of the studies, respectively. However, they played crucial roles in specific tasks. Self-presence, prominent in sensory and multisensory interaction tasks and rehabilitation and bodily tasks, improved the sense of identity of users, increasing the participation in tasks \cite{koulouris2020me}\cite{michael2020race}\cite{wolf2021embodiment}\cite{mal2023impact}. Unruh et al.'s study \cite{unruh2023body} found that high levels of self-presence influenced users' perception of time, deepening their immersive experience.

\subsubsection{Task Types}

\begin{figure}[ht]
    \centering
    \includegraphics[width=14cm]{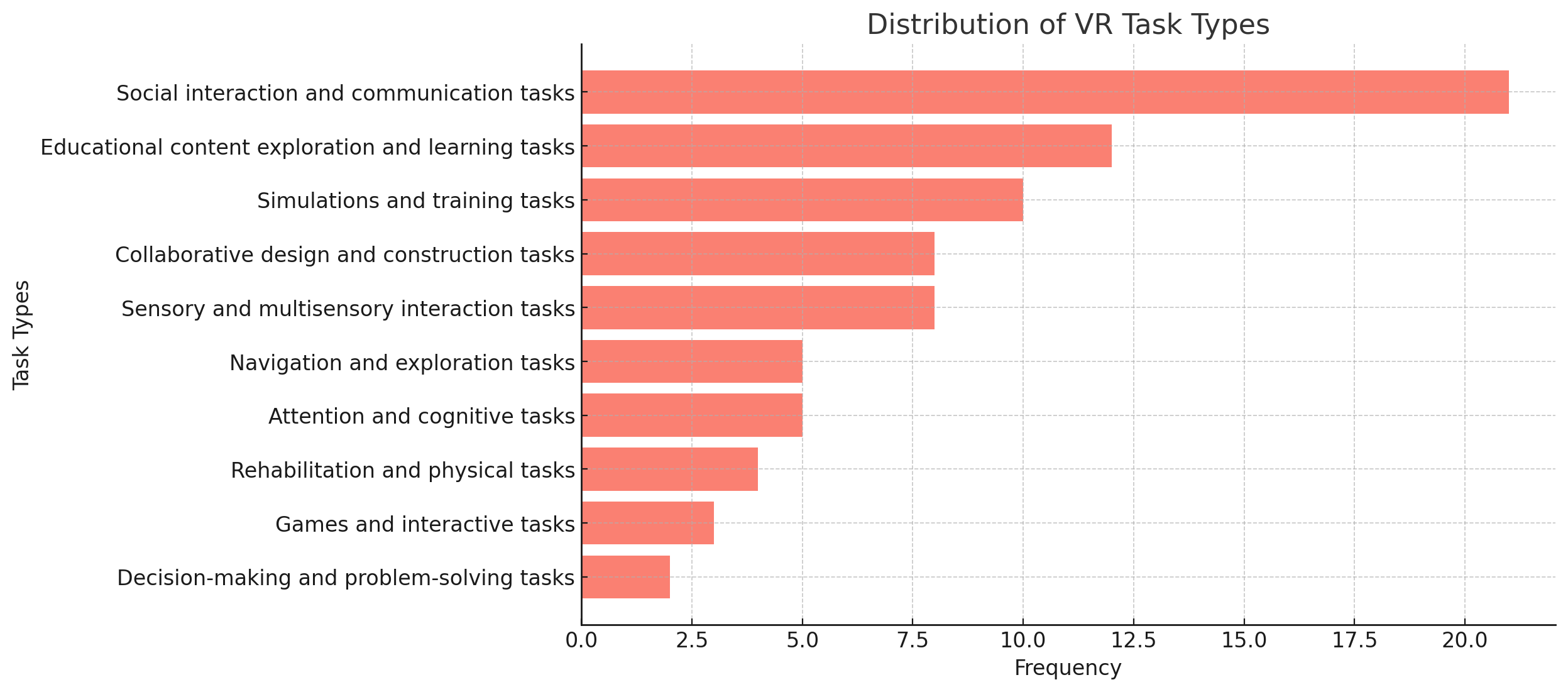}
    \caption{Statistics on the task types used in the studies.}
    \label{task- type}
\end{figure}

In terms of task types, as shown in Figure \ref{presencetype} (2). ``Social Interaction and Communication Tasks'' are the most common category in research, accounting for 27\%. These tasks typically involve users collaborating and interacting through virtual environments to enhance social connections. For example, the study by Gu et al. \cite{gu2022role} allows participants to experience the roles of bully and victim through role swapping, thereby enhancing understanding and empathy towards the issue of school bullying, as shown in figure \ref{TaskTypes} (a). Secondly, ``Educational Content Exploration and Learning Tasks'' account for 15\%. These tasks emphasize interactive learning and immersive experiences in virtual environments. For instance, Bellgardt et al. \cite{bellgardt2023virtual} reimplemented the virtual optical bench in a VR environment, enhancing students' understanding of optical systems through an immersive and accurate VR simulation. ``Simulations and Training Tasks'' come next, accounting for 13\%. These tasks highlight the user's ability to learn and train skills in virtual environments. For instance, Li et al. \cite{li2020analysing} developed a virtual reality operating room simulator for surgeons to train in laparoscopic surgery, as shown in Figure \ref{TaskTypes} (c). Through a highly realistic simulation environment, doctors can enhance their skills without any risk. The next is ``Collaborative Design and Construction Tasks'' and ``Sensory and Multisensory Interaction Tasks'', each accounting for 10\%. The former emphasize teamwork and problem-solving in virtual environments. For example, Tian et al. \cite{tian2023using} found that virtual replicas improved the efficiency of MR remote collaboration by improving the understanding of remote instructions during an object assembly task, outperforming the 3D annotation drawing. The latter emphasizes multisensory interactions in VR, as shown in figure \ref{TaskTypes} (d). For example, Yang et al. \cite{yang2022hybridtrak} presented HybridTrak, a system that enhances full-body tracking in VR by using a single RGB webcam in combination with upper body inside-out tracking, offering more accurate and natural body postures compared to traditional RGB or depth-based methods, as shown in Figure \ref{TaskTypes} (e). Other types of tasks, such as ``Navigation and Exploration Tasks'' and ``Attention and Cognitive Tasks'' each account for 6\%, ``Rehabilitation and Physical Tasks'' (5\%), ``Gaming and Interactive Tasks'' (4\%), and ``Decision-Making and Problem-Solving Tasks'' (3\%).

\begin{figure}[ht]
    \centering
    \includegraphics[width=10cm]{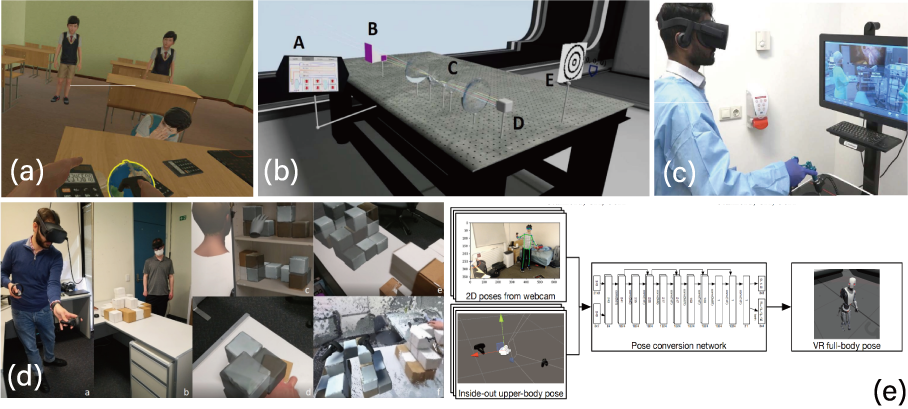}
    \caption{In terms of task types: (a) Social Interaction and Communication Tasks \cite{gu2022role}; (b) Educational Content Exploration and Learning Tasks \cite{bellgardt2023virtual}; (c) Simulations and Training Tasks \cite{li2020analysing}; (d) Collaborative Design and Construction Tasks \cite{tian2023using}; (e) Sensory and Multisensory Interaction Tasks \cite{yang2022hybridtrak}.}
    \label{Task Types}
\end{figure}

\subsection{Interactions Between Presence Types in VR Tasks}

\subsubsection{Intersections of Presence Types in VR}
We found that many studies involve multiple types of presence. The intersections of these presence types reveal the complexity and multidimensionality of the user experience in VR environments. Through data analysis, we identified the following main combinations of presence types and their applications in different tasks. 

First, the intersection of social presence \& co-presence is the most extensive, with a total of 27 studies. This combination highlights the importance of social interaction and shared experience in multi-user environments. For instance, Yoon et al.'s study \cite{yoon2020evaluating} explored the role of remote virtual hand models in gesture collaboration tasks, finding that more realistic hand models significantly improved users' social presence and task performance. This suggests that the combination of social and co-presence can enhance collaborative cooperation and information sharing among users, especially in task scenarios requiring rich interaction.

Second, the intersection of spatial presence \& cognitive presence has reached 23 studies. This combination emphasizes the importance of spatial perception for user cognitive engagement. For example, Wijayanto et al.\cite{wijayanto2023comparing} found that even in VR with non-realistic rendering, participants could accurately perceive the size of targets through invariant information in the environment. This indicates that spatial presence not only enhances the immersive experience, but also facilitates deep cognition and perceptual adaptation.

Third, the intersection of spatial presence \& social presence has 17 studies. This combination suggests that researchers emphasize the integration of spatial perception and social interaction when designing virtual environments. For example, Schott et al.\cite{schott2024excuse} proposed improving users' obstructed views in virtual museums through transparency and horizontal displacement techniques, enhancing collaboration and social satisfaction, highlighting the importance of spatial presence in optimizing virtual environment experiences. Other presence types, such as ``Cognitive presence and self-presence'' have nine studies, ``Spatial Presence \& Co-presence'' have three studies, ``Social Presence \& Co-presence'' have two studies, and ``Spatial Presence \& Cognitive Presence \&  Co-Presence'' has one study.


\subsubsection{Mapping Presence Types to Task Functions}

\begin{figure}[ht]
    \centering
    \includegraphics[width=13cm]{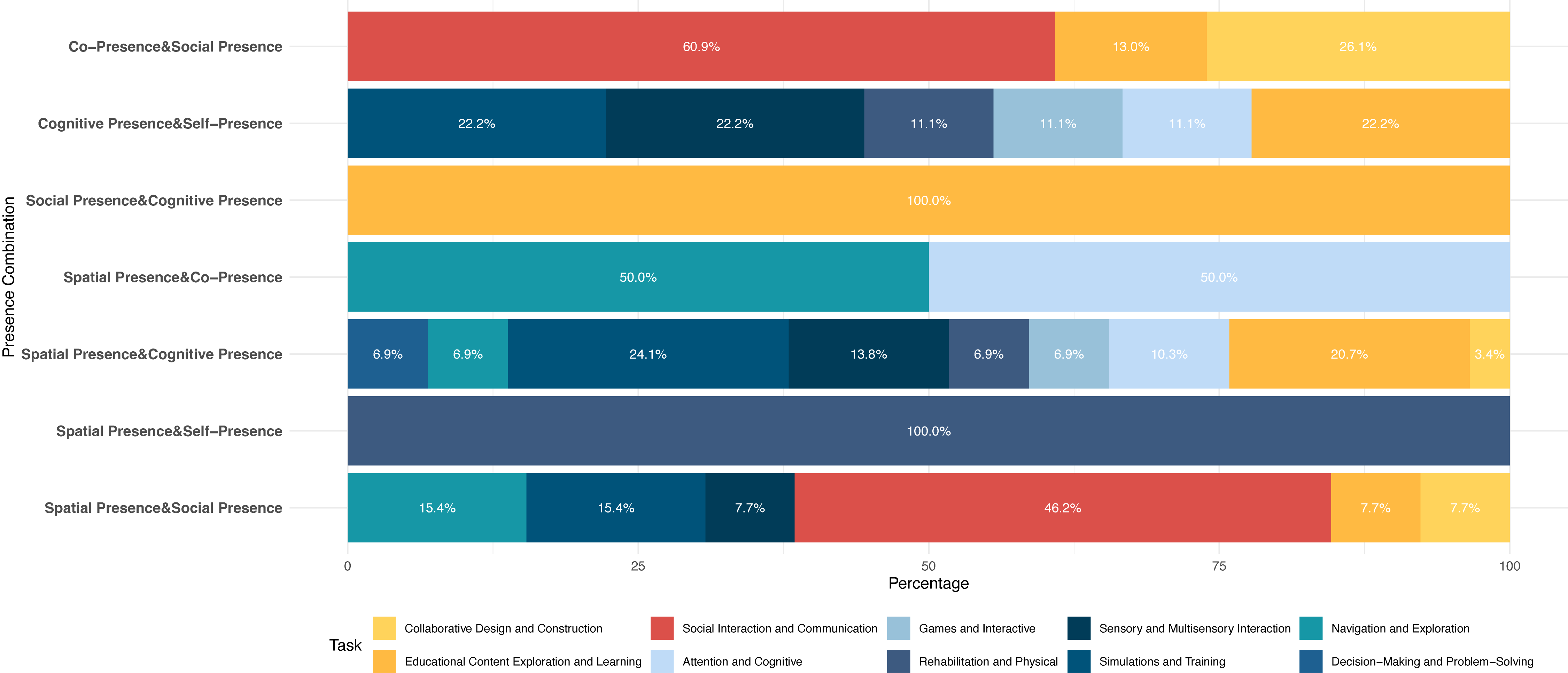}
    \caption{Stacked column chart of the percentages of different presence types and corresponding task types.}
    \label{Mutil-presence type}
\end{figure}
We observed potential associations between different types of presence and task categories, as illustrated in Figure \ref{Mutil-presence type}. Spatial presence plays a foundational role in tasks that require users to navigate, explore, or simulate real-world scenarios within a virtual environment. When considered independently, it is primarily associated with self-presence tasks (100.0\%), highlighting its essential role in fostering an immersive sense of self in virtual environments.

When spatial presence is combined with other types of presence—such as spatial + cognitive or spatial + social—its functional scope expands to support more complex and multifaceted tasks. For instance, the combination of spatial + social presence accounts for the largest proportion (46.2\%), indicating that it significantly contributes to social engagement and collaborative virtual interactions. Similarly, spatial + cognitive presence is distributed across multiple task types, with 24.1\% for simulations and training and other smaller percentages (6.9\%–20.7\%) across additional categories, reflecting its importance in supporting diverse cognitive processes in virtual spaces.

Cognitive presence is indispensable for facilitating deep learning, critical thinking, and problem-solving. When it stands alone, it is distributed across several tasks, with the highest percentage being 22.2\% in two categories and smaller proportions (11.1\%) in others. This underscores its role in fostering critical engagement and conceptual understanding. When combined with spatial presence, cognitive presence is particularly impactful, contributing to 24.1\% of spatial-cognitive tasks, suggesting a synergy that enhances users' ability to process and internalize complex information.
Social presence, when it exists independently, primarily serves cognitive presence-related tasks (100.0\%), highlighting its crucial role in supporting social learning and interactive engagement. When combined with co-presence, social presence contributes notably to 60.9\% of social interaction and communication, reinforcing its significance in tasks requiring shared understanding and teamwork. The presence of co-presence alone is not explicitly represented as an independent category, but when combined with spatial presence, it is equally split (50.0\%–50.0\%) between two key categories (navigation and exploration, attention and cognitive), reflecting its dual importance in immersive and collaborative virtual experiences.

Although co-presence appears less frequently in studies, it exhibits significant roles in specific tasks that demand high levels of collaboration. The combination of co-presence + social presence accounts for 60.9\% in social interaction and communication, illustrating its importance in promoting mutual awareness and effective teamwork.

In summary, different presence designs have unique functions in specific task scenarios:

(1) Spatial presence serves as the foundation for achieving immersive virtual reality experiences, providing users with realistic spatial awareness and the ability to interact naturally within virtual environments.

(2) Cognitive presence is central to promoting knowledge acquisition, deep understanding, and critical thinking. It enables users to engage meaningfully with content, reflect on their experiences, and apply new insights effectively.

(3) Social presence and co-presence play key roles in team collaboration, effective communication, and the development of social connections. They are essential for tasks that require cooperation, shared understanding, and collective problem-solving.

By understanding these associations, designers and researchers can tailor virtual reality environments to meet the nuanced needs of various educational, training, and collaborative tasks. This targeted approach can enhance the user experience by providing the appropriate type of presence needed for specific tasks, thereby improving task performance efficiency, learning outcomes, and overall satisfaction. Emphasizing the interplay between different types of presence allows for the creation of more engaging, effective, and meaningful virtual experiences that align closely with users' objectives and the demands of complex tasks.

\subsection{Evaluation of Tasks}
We examined the evaluation methods used in the included studies, noting both objective performance metrics and subjective user reports. Overall, 44.9\% (35/78) of the papers adopted a single assessment approach—often self-report questionnaires—while 55.1\% (43/78) integrated multiple methods. Among these, 75  papers employed quantitative evaluations, such as measuring task performance (e.g., completion time, accuracy) and administering standardized questionnaires. Commonly used instruments included the IGroup Presence Questionnaire (IPQ) \cite{melo2023much}, NASA Task Load Index (NASA-TLX) \cite{hart2006nasa}, Simulator Sickness Questionnaire (SSQ), Virtual Embodiment Questionnaire (VEQ), and the System Usability Scale (SUS). Meanwhile, 3 papers relied primarily on qualitative methods, such as semi-structured interviews and thematic analyses. In addition, approximately 8\% of the studies collected physiological data (e.g., heart rate, eye-tracking, EEG) to objectively assess user states \cite{sasikumar2024user,michael2020race}. Notably, several papers combined self-reports, task performance metrics, and physiological measures to enhance the reliability and depth of their findings. For instance, Sasikumar et al. \cite{sasikumar2024user} showed a positive correlation between co-presence and task performance by integrating self-report scales, objective performance data, and physiological recordings. Michael et al. \cite{michael2020race} demonstrated a high correlation between self-competition and intrinsic motivation, underscoring the advantages of blending objective metrics with self-report data.

\subsection{Presences Enhance Collaboration and Work Efficiency in VR}

\subsubsection{Enhance Presence and collaboration efficiency}

\begin{figure}[ht]
    \centering
    \includegraphics[width=11cm]{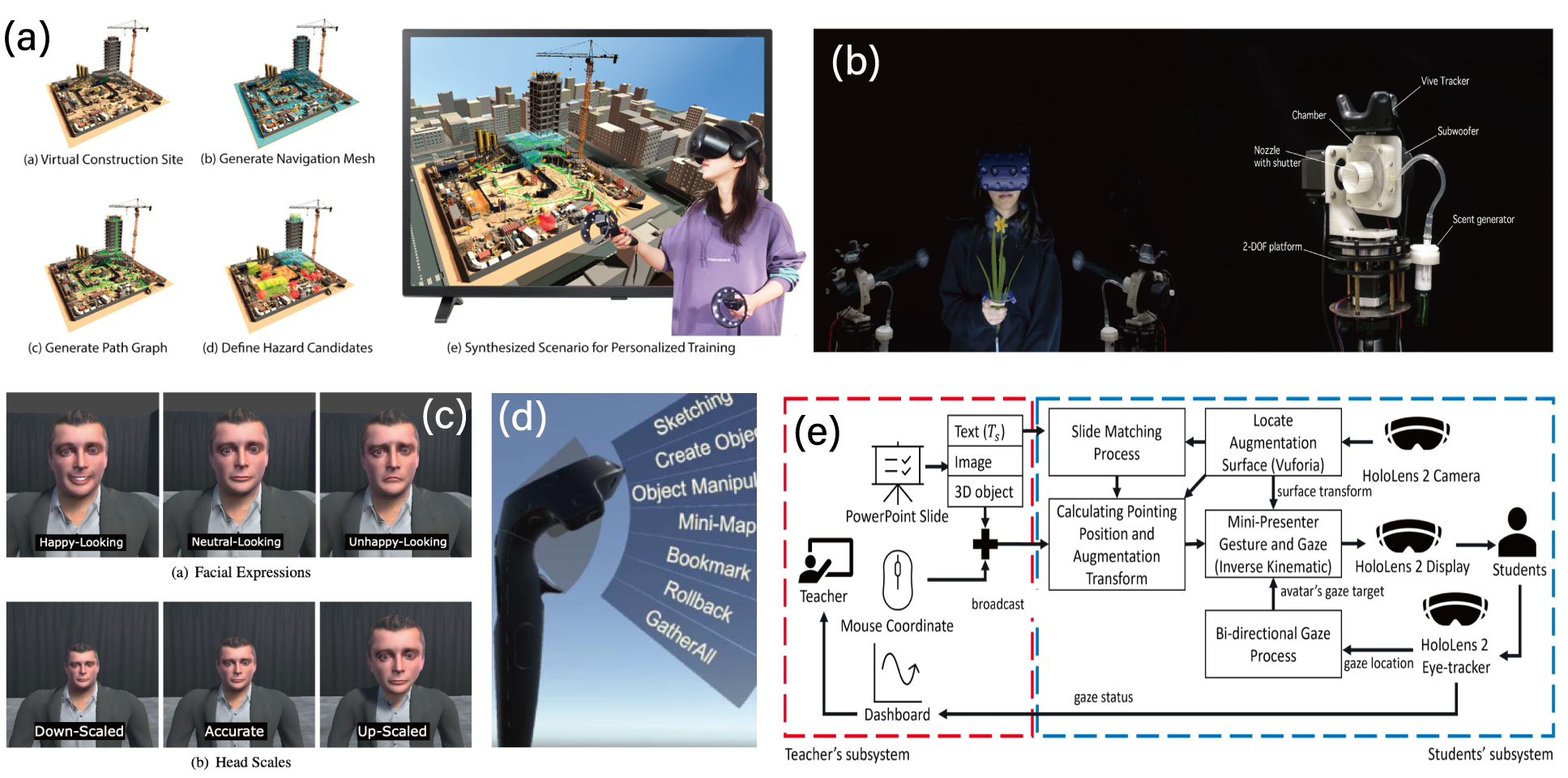}
    \caption{Five primary design methods: (a) Environment Design \cite{li2022synthesizing}; (b) Multimodal Sensory Design \cite{hu2021abio}; (c) Avatar Design \cite{choudhary2023exploring}; (d) GUI Interface Design \cite{hsu2020design}; (e) Biometric Data Integration \cite{thanyadit2023tutor}.} 
    \label{Enhance}
\end{figure}

Through a systematic review of 78 relevant studies, five primary design methods were identified as critical for enhancing presence in VR environments, as shown in Figure \ref{Enhance}. These methods include environmental design (89.7\%), multimodal sensory design (57.7\%), Avatar design (57.7\%), GUI interface design (29.5\%), and biometric data integration (7.7\%). 

\textbf{Environmental Design.}
Among the 78 reviewed studies, 70 studies emphasized environment design as a foundational method for enhancing presence in VR environments. These studies consistently demonstrate that environment design not only serves as the basis for constructing immersive virtual spaces but also directly influences task performance and user interaction within these spaces. Three prominent aspects emerge within this category: high-fidelity scene rendering, dynamic scene interaction, and adaptive environmental feedback.

High-fidelity scene rendering remains a cornerstone of presence enhancement, relying on advanced techniques such as real-time ray tracing, texture mapping, and dynamic shadow effects. These techniques enable virtual environments to replicate real-world visual characteristics with a high degree of precision. For example, the virtual optical workstation utilizes real-time ray tracing technology to simulate the interaction between laser beams and optical lenses, enabling users to directly observe the changes in light paths. This enhances their intuitive understanding of optical principles and their sense of immersion \cite{bellgardt2023virtual}.

In addition to static visual fidelity, dynamic scene interaction plays a crucial role in optimizing presence. Unlike static environments, dynamic designs allow virtual elements to adapt in real-time based on user actions and system feedback \cite{thanyadit2023tutor,li2022synthesizing}. In collaborative VR classrooms, for example, the system dynamically adjusts the layout of visual elements based on students' gaze patterns, minimizing visual clutter and ensuring the visibility of critical content \cite{bozkir2021exploiting}. These mechanisms enable VR environments to respond intelligently to users' states and actions, ensuring a fluid and intuitive interactive experience.

\textbf{Multimodal Sensory Design.} 
Multimodal sensory design was explored in 45 studies, underscoring its importance in reducing sensory dissonance and enhancing immersion. Multimodal design involves the integration of visual, auditory, haptic, and occasionally olfactory feedback, creating a cohesive and synchronized sensory experience.

Visual feedback remains the central dimension of multimodal design. High-resolution textures, dynamic visual cues, and real-time object feedback were frequently employed to enhance spatial perception and environmental coherence. For example, in virtual navigation tasks, dynamic path markers guide users’ focus toward essential objectives, effectively reducing cognitive load \cite{li2022synthesizing}. Auditory feedback, particularly through spatial audio and dynamic sound source localization, ensures that users can perceive sound direction and distance naturally. In collaborative scenarios, systems dynamically adjust the volume and direction of speakers' voices, ensuring clarity and minimizing disruptions during conversations \cite{lee2024may}. Haptic feedback introduces a tactile dimension to virtual interaction, achieved through vibration cues, force feedback devices, and pressure-sensitive inputs. In virtual manipulation tasks, for instance, haptic devices allow users to feel the resistance and weight of virtual objects, improving realism and interaction precision \cite{yun2024exploring, michael2020race}. Though less common, olfactory feedback has also been explored. For example, in virtual garden simulations, scent generators synchronize with visual and tactile cues, releasing floral fragrances when users interact with virtual flowers \cite{hu2021abio}. Such integration further deepens sensory immersion and emotional resonance.

\textbf{Avatar Design.}
Avatar design was examined in 45 studies, where Avatars function as both identity representations and emotional communication conduits in VR environments. Two primary themes emerged from these studies: high-fidelity anthropomorphic design and customization for self-expression. High-fidelity anthropomorphic design focuses on replicating realistic facial expressions, body movements, and mirror reflections. Advanced facial tracking and gesture synchronization technologies ensure that Avatars accurately represent users' emotional states and physical gestures. For example, in collaborative VR tasks, high-fidelity Avatars facilitate non-verbal emotional exchanges, reducing misunderstandings during team interactions \cite{choudhary2023exploring,kimmel2023let}. Customization for self-expression allows users to personalize their avatars’ appearance, clothing, and gestural behaviors of their avatars, reflecting their preferences or self-identity \cite{sykownik2022something}. Such design flexibility not only enhances user engagement but also fosters a stronger sense of ownership and social trust.

\textbf{GUI Interface Design.}
23 studies explored the role of Graphical User Interface (GUI) design in enhancing presence and interaction efficiency in VR environments. GUI design primarily revolves around dynamic visual prompts \cite{reichherzer2022supporting}, real-time information feedback \cite{sasikumar2024user}, and interactive control interfaces \cite{hsu2020design}. These elements serve to bridge the gap between users and the virtual environment, enabling more intuitive and responsive interactions.


\textbf{Biometric Data Integration.}
Biometric data integration was discussed in 6 studies. Key biometric indicators include heart rate, cognitive load, and attention levels, which are often monitored using technologies such as EEG (electroencephalography) and eye-tracking systems \cite{10.1145/3492802,voigt2021don}. These indicators provide a deeper layer of feedback by offering real-time insights into users' emotional and cognitive conditions. 

\subsubsection{The synergistic effect of multimodal feedback}

\begin{figure}[ht]
    \centering
    \includegraphics[width=11cm]{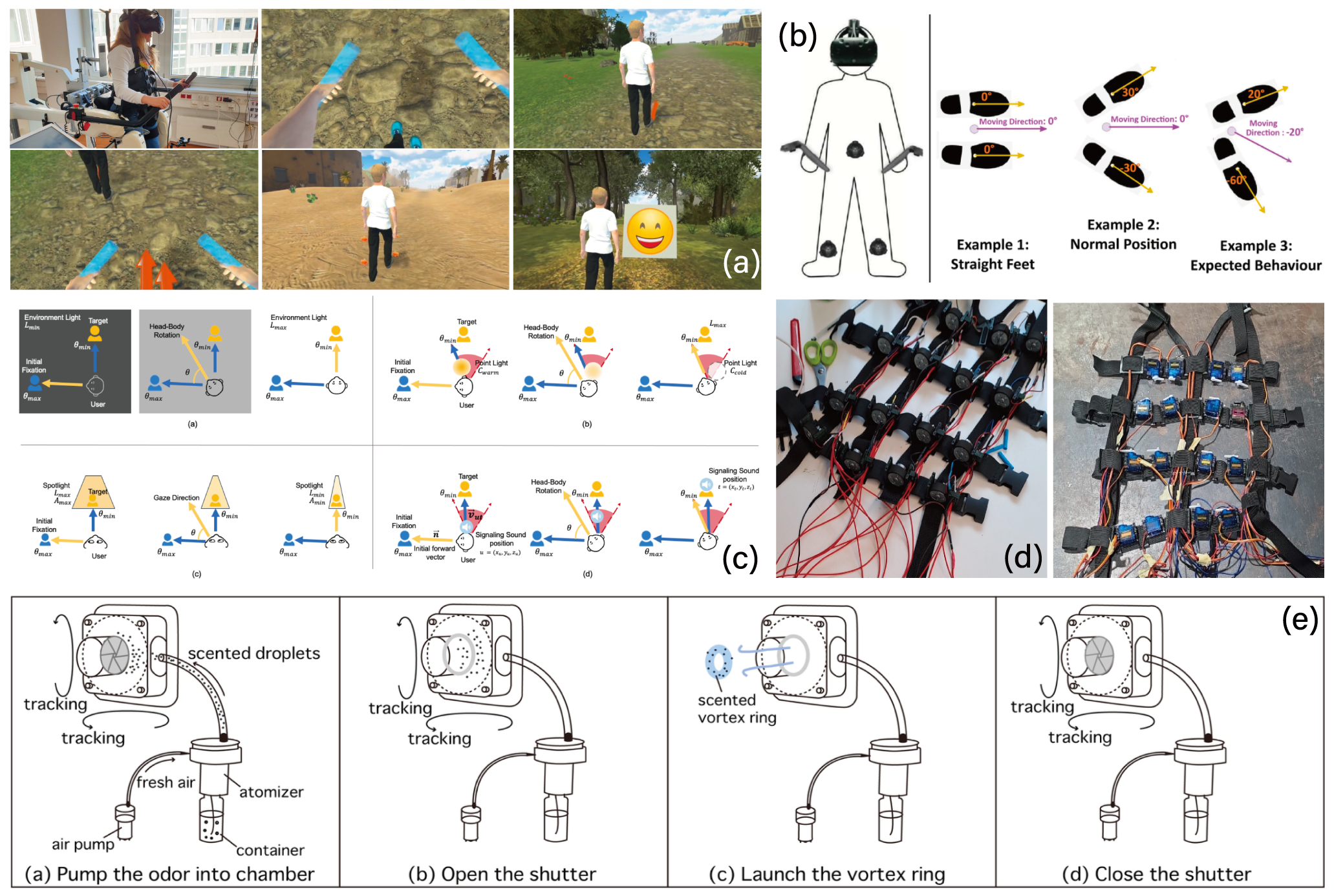}
    \caption{Five primary sensory dimensions of learner perception: (a) Visual Perception \cite{hamzeheinejad2021impact}; (b) Kinesthetic Perception \cite{10.1145/3641825.3687735}; (c) Auditory Perception \cite{lee2024may}; (d) Tactile Perception \cite{espositi2024room}; (e) Olfactory Perception \cite{hu2021abio}.} 
    \label{synergistic}
\end{figure}

In a systematic review of relevant studies, five primary sensory dimensions of learner perception were identified: visual perception (98.7\%), auditory perception (39.7\%), kinesthetic perception (29.5\%), tactile perception (19.2\%), and olfactory perception (1.3\%), as shown in Figure \ref{synergistic}.

Visual perception emerges as the most frequently discussed sensory dimension, with nearly all studies emphasizing its critical role in information acquisition, attention maintenance, and task execution efficiency. The primary focus of visual perception lies in the precise delivery and visualization of information. Techniques such as visual highlighting of key task elements and attention-guiding mechanisms are widely employed to help learners efficiently identify and focus on essential information \cite{bellgardt2023virtual}. Additionally, real-time data visualization enables learners to instantly access system responses and feedback, enhancing the timeliness and accuracy of their decisions \cite{michael2020race}. Multi-perspective switching mechanisms further provide learners with flexible observation angles, allowing them to seamlessly transition between global overviews and localized details based on specific task requirements \cite{moss2024going,wei2023feeling}. These strategies not only ensure clarity and transparency in information delivery but also offer sustained visual support throughout different stages of task execution, ultimately enhancing learners' sense of presence and engagement.

Auditory perception is addressed in 31 studies, with particular emphasis on its role in spatial awareness, cognitive load reduction, and emotional resonance. Spatial audio technologies, such as 3D directional sound effects, are central to auditory perception, enabling learners to accurately locate sound sources and improve spatial orientation during tasks. Real-time auditory feedback, including alert sounds and task-completion chimes, is equally significant in providing immediate status updates, reducing response time, and enhancing situational awareness \cite{lee2024may}. Additionally, auditory feedback serves an emotional regulatory function, with background sounds helping to create calming environments and alleviate anxiety during high-pressure tasks \cite{wang2023designing}. Voice guidance systems play an essential role in clarifying task instructions, reducing ambiguity, and streamlining communication in complex task scenarios.

Kinesthetic perception is examined in 23 studies, focusing on spatial interaction, motor skill acquisition, and embodied learning experiences. Technologies such as full-body motion tracking and gesture-based interaction enable learners to interact intuitively with virtual environments through natural physical movements, including grabbing, pointing, and rotating \cite{shigyo2024vr}. These interactions enhance both the physical realism and spatial coordination of tasks. Motion synchronization technologies help minimize sensory discrepancies and optimize interaction fluidity. For example, accurate gait tracking allows learners to adjust their movements based on visual cues, improving precision in task execution \cite{hamzeheinejad2021impact}. Furthermore, kinesthetic perception supports spatial awareness, helping learners navigate complex virtual environments with efficiency and confidence.

Tactile perception is discussed in 15 studies, primarily addressing physical interaction fidelity, tactile feedback mechanisms, and material property simulation. Vibration feedback systems, delivered through handheld controllers or wearable devices, provide immediate tactile sensations such as impact, texture, and resistance, enhancing the realism of physical interactions \cite{espositi2024room,gomi2024inflatablebots}. Material properties, including weight, texture, and flexibility, are also an important point, particularly in scenarios requiring fine motor control, such as virtual surgical training \cite{li2020analysing}. Force feedback mechanisms simulate physical resistance during operations (e.g., tightening screws or pulling levers), improving the accuracy and authenticity of task performance. 

Olfactory perception is mentioned in one study. Scent feedback can deepen learners' emotional connection to specific scenarios, thereby enhancing their overall sense of presence. For example, in a virtual natural environment, the release of floral scents upon touching a virtual flower creates a multisensory experience that reinforces emotional engagement and memory retention \cite{hu2021abio}.

\subsubsection{The relationship between presence and cognitive load}
In our analysis, 36 studies examine how different types of presence affect cognitive load, impacting learning efficiency, task performance, and overall user experience in various VR settings. Of these studies, 72\% (26/36) focus on spatial and cognitive presence, exploring how the sensation of being physically situated within a virtual environment affects users' ability to process information, perform tasks, and retain knowledge. For instance, Wijayanto et al.\cite{wijayanto2023comparing} compared visual realism in VR versus real-world viewing and reported significant positive correlations, indicating that higher spatial presence facilitated better size perception and spatial judgments. Conversely, 25\% (9/36) of the studies emphasize self and cognitive presence, examining how interactions with virtual humans or other users modulate cognitive load. For example, Gu et al. \cite{gu2022role} proposed a novel role-exchange playing paradigm, where users play two opposite roles successively in immersive virtual environments, finding that it enhances the moral understanding, empathy, and commitment to stopping bullying among students, potentially extending to applications in counseling, therapy, and crime prevention. Based on the analysis of these studies, the relationship between presence and cognitive load can be categorized into four main impacts:

\begin{enumerate}
    \item \textbf{Enhanced Learning Efficiency:} Spatial presence significantly improves cognitive task performance by providing a more immersive and contextually rich environment, thereby facilitating better information processing and retention. For example, Reichherzer et al.\cite{reichherzer2022supporting} found that VR improved the spatial comprehension of forensic evidence, enhancing jury understanding in virtual courtroom simulations.
    
   \item \textbf{Optimized Cognitive Resource Allocation:} Social and Cognitive Presence in collaborative VR tasks help distribute the cognitive load more effectively, enhancing individual learning while fostering collective problem-solving skills. For example, Petersen et al. \cite{petersen2021pedagogical} explored the impact of pedagogical agents in VR on learning outcomes, finding that the appearance and behavior of a virtual museum guide influenced factual and conceptual knowledge acquisition, offering new insights into the use of agents in immersive education.
    
    \item \textbf{Cognitive Flexibility and Adaptation:} High levels of presence enable users to adapt more readily to complex tasks and dynamic environments, promoting cognitive flexibility and resilience in skill acquisition. For instance, Wijayanto et al.\cite{wijayanto2023comparing} demonstrated that higher visual realism in VR environments enhanced spatial awareness and reduced cognitive load during size perception tasks.
    
    \item \textbf{Mitigation of Cognitive Overload:} While presence generally aids cognitive load management, some studies highlight potential drawbacks, such as increased frustration or motion sickness under specific conditions. Bueno-Vesga et al.\cite{bueno2021effects} found that higher presence was associated with reduced frustration, indicating that certain presence-enhancing features can mitigate negative cognitive states.
\end{enumerate}

The analysis of these 36 studies illustrates that presence is intricately linked to cognitive load in VR environments, significantly influencing learning efficiency, task performance, and overall user experience. Spatial presence enhances cognitive processing and information retention by immersing users in contextually rich virtual settings, while social and cognitive presences facilitate effective cognitive resource allocation through collaborative interactions. Additionally, cognitive presence fosters deep mental engagement and flexibility in problem-solving tasks. However, the relationship between presence and cognitive load is nuanced, with certain presence-enhancing elements potentially increasing cognitive strain under specific conditions. Therefore, designing VR experiences that strategically balance immersive presence with cognitive load demands is crucial for optimizing skill development and ensuring user well-being.

\subsection{Exercise and skill training}

\subsubsection{Impact of social presence on group training}

Group-based VR training studies frequently measure and discuss social presence through questionnaires (e.g., Social Presence Questionnaire, Networked Minds), semi-structured interviews, and objective performance metrics (e.g., task completion time and communication quality). Elevated social presence often enhances learning motivation and collaboration engagement: for instance, 
Sasikumar et al. \cite{sasikumar2024user} indicate that heightened co-presence during a collaborative engine assembly activity is associated with expedited task completion, enhanced empathy, and intensified social bonding, 
while Hsu et al. \cite{hsu2020design} show that near-real-time remote visualization and audio interaction in a VR architectural design environment lead to greater user involvement and more efficient collaboration. 

Overall, higher social presence tends to align—though not always strictly—with improved learning outcomes and team performance \cite{schott2023vreal,lee2024may,thanyadit2023tutor,jin2023collaborative,tian2023using,abramczuk2023meet}, yet in large-scale multi-user settings, it can sometimes distract from the task or increase ``social pressure.'' For example, in Jin et al.’s \cite{jin2023collaborative} collaborative video-learning scenario, the impact of social presence on test-score gains was overshadowed by its effect on user satisfaction and group rapport. Multiple studies also show that adding richer input modalities—such as motion capture, facial tracking, or physiological sensing—can boost the sense of ``being there'' and mutual trust, but at the cost of higher hardware and technical complexity \cite{sasikumar2024user,kimmel2023let}. Beyond skill-based tasks, social presence supports emotional well-being and peer learning: Gu et al.’s \cite{gu2022role} anti-bullying role-play approach, Wei et al.’s \cite{wei2023feeling} creative lighting design collaboration, and Petersen et al.’s \cite{petersen2021pedagogical} museum-based educational agents all indicate that more substantial social presence reinforces group cohesion, sparks emotional resonance, and fosters learning or attitudinal shifts. 

\begin{figure}[ht]
    \centering
    \includegraphics[width=13cm]{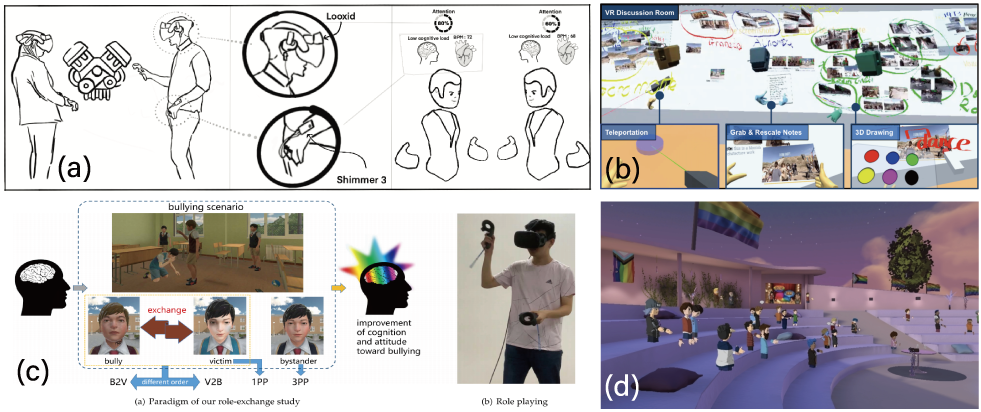}
    \caption{Four Types of Multi-User Training or Learning Studies Based on Social Presence: (a) remote expert + local operator with shared viewpoints and tool access \cite{sasikumar2024user}; (b) distributed learners concurrently engaging in discussion-based tasks \cite{jin2023collaborative}; (c) role-play or immersive moral education with multiple users \cite{gu2022role}; and (d) community-focused instruction on large-scale social VR platforms \cite{li2023we}.}
    \label{social presence}
\end{figure}

Broadly, these multi-user training or learning studies span four key paradigms, as shown in Figure\ref{socialpresence}. (a) remote expert + local operator with shared viewpoints and tool access \cite{sasikumar2024user,hsu2020design}, (b) distributed learners concurrently engaging in discussion-based tasks \cite{schott2021vr,jin2023collaborative,petersen2021pedagogical}, (c) role-play or immersive moral education with multiple users \cite{gu2022role,lee2024may,wei2023feeling}, and (d) community-focused instruction on large-scale social VR platforms \cite{sykownik2022something,li2023we,wei2023bridging}. These lines of research collectively suggest that where tasks demand strong collaboration, emotional investment, or attitude change—such as remote co-design, VR surgical training, or immersive multi-role learning—greater social presence usually boosts team cohesion, learner engagement, empathy, and motivation, thereby improving performance.

\subsubsection{The influence of self-presence on motor control and sense of reality}
An increasing body of research underscores the significance of self presence, defined as one’s subjective sense of embodiment and alignment with one’s virtual self, in shaping motor control and perceived realism in VR experiences. Among the 78 studies surveyed, ten focus specifically on self-presence. Of these, 40\% (4/10) concentrate on physical/motion-oriented tasks requiring explicit motor control, such as exergames or full-body interactions. For example, Koulouris et al. \cite{koulouris2020me} and Alexander Michael \cite{michael2020race} investigated VR games where closely matched real and virtual movements not only increase users’ immersion but also boost motivation and performance. Similarly, Mal et al. \cite{mal2023impact} explore how avatar-environment congruence affects plausibility and embodiment in rehabilitation or physical tasks, while Bhargava et al. \cite{bhargava2023empirically} examine how manipulating eye height and self-avatars influences dynamic passability.

The remaining 60\% (6/10) of these studies emphasize emotional or conceptual experiences, focusing on the experiential or reflective aspects of self-presence. These include Shigyo et al.’s \cite{shigyo2024vr} VR-mediated cognitive defusion for managing negative thoughts, Jicol et al.’s \cite{jicol2021effects} exploration of how emotional states (e.g., fear or joy) and agency impact presence, and Sojung Bahng et al.’s \cite{bahng2020reflexive} reflexive VR storytelling aimed at fostering self-reflection on death and loneliness. Gu et al. \cite{gu2022role} employ role-exchange play in anti-bullying scenarios to investigate empathy and moral education, Wolf et al. \cite{wolf2021embodiment} evaluate how photorealistic avatars influence body-weight perception and body image, and Unruh et al. \cite{unruh2023body} examine how varying degrees of virtual embodiment alter users’ sense of time.

Collectively, these studies often link self-presence to performance, movement, and subjective experience. For instance, stronger alignment between real-world and virtual-body movements typically yields enhanced immersion and motivation, although overly intense motion cues can lead to increased motion sickness or discomfort. Conversely, in more emotionally or conceptually driven scenarios, self-presence can promote deeper empathy, introspection, or emotional engagement, sometimes at the cost of greater psychological discomfort. Overall, these findings highlight the pivotal role of self-presence in bridging users’ physical actions with the virtual environment, thus creating compelling VR experiences. Whether in physically demanding tasks or emotionally rich narratives, striking the right balance between technical fidelity, task design, and user comfort is crucial to maximizing the benefits of self-presence across diverse virtual reality applications.

\subsubsection{The role of multisensory feedback in skill acquisition}

Among the 78 reviewed studies, 14 specifically focus on skill acquisition, categorized by academic discipline and learning mode. Academically, these studies fall into four domains: Medical and Health Sciences (28.6\%), Engineering and Technical Sciences (14.3\%), Arts and Media (14.3\%), and Social and Behavioral Sciences (42.9\%). In terms of learning modes, they are classified into individual learning (50\%), collaborative learning (35.7\%), and a combination of both (14.3\%). Multisensory feedback plays a central role in facilitating skill acquisition, enhancing learners’ ability to intuitively grasp complex information and execute precise tasks. However, the application and emphasis of multisensory feedback exhibit notable differences across disciplines and learning modes.

In different academic disciplines, multisensory feedback serves distinct purposes shaped by the unique demands of each field. The Medical and Health Sciences prioritize precision and transferability in the application of multisensory feedback. In contexts such as virtual surgery training and motor rehabilitation, tactile feedback simulates tool resistance, auditory cues deliver real-time instructions, and visual feedback illustrates dynamic physiological changes \cite{hamzeheinejad2021impact,li2020analysing}. These mechanisms ensure learners can perform refined tasks and make timely decisions in highly immersive, realistic environments. In Engineering and Arts and Media, multisensory feedback serves distinct yet complementary purposes: parameter adjustment and creative expression \cite{wei2023feeling,wei2024hearing}. For example, in optical path calibration, tactile feedback reinforces realism through fine-tuned equipment resistance, while in sound recording, auditory cues guide learners in identifying spatial audio balance and source directionality. In contrast, Social and Behavioral Sciences emphasize attention management and social interaction. Through eye-tracking mechanisms, learners’ focal points are monitored, while virtual avatars transmit non-verbal cues to strengthen social presence and facilitate collaborative engagement \cite{thanyadit2023tutor,jin2023collaborative}.

From a learning mode perspective, individual learning leverages multisensory feedback to enhance autonomy and precision. In surgical training or optical parameter calibration, tactile cues allow learners to perceive minute adjustments, visual cues provide dynamic, real-time insights, and auditory cues deliver immediate alerts or confirmations. This closed-loop feedback system enables focused, iterative practice, fostering deep skill mastery in controlled virtual settings \cite{wei2024hearing,hamzeheinejad2021impact}. In contrast, collaborative learning emphasizes information sharing and role coordination \cite{jin2023collaborative,bozkir2021exploiting}. For example, in collaborative anatomical explorations, visual cues ensure shared reference points via synchronized 3D models, auditory cues facilitate seamless communication through spatial audio, and avatars assist in conveying nuanced gestures and expressions. This integrated multisensory feedback framework optimizes collaborative efficiency and reinforces collective task achievement. The hybrid learning mode combines the strengths of both approaches, balancing the autonomy and precision of individual learning with the coordinated efficiency of team-based interaction, demonstrating adaptability across varied learning contexts.

\section{DISCUSSION}\label{section4}
\begin{figure}[ht]
    \centering
    \includegraphics[width=14cm]{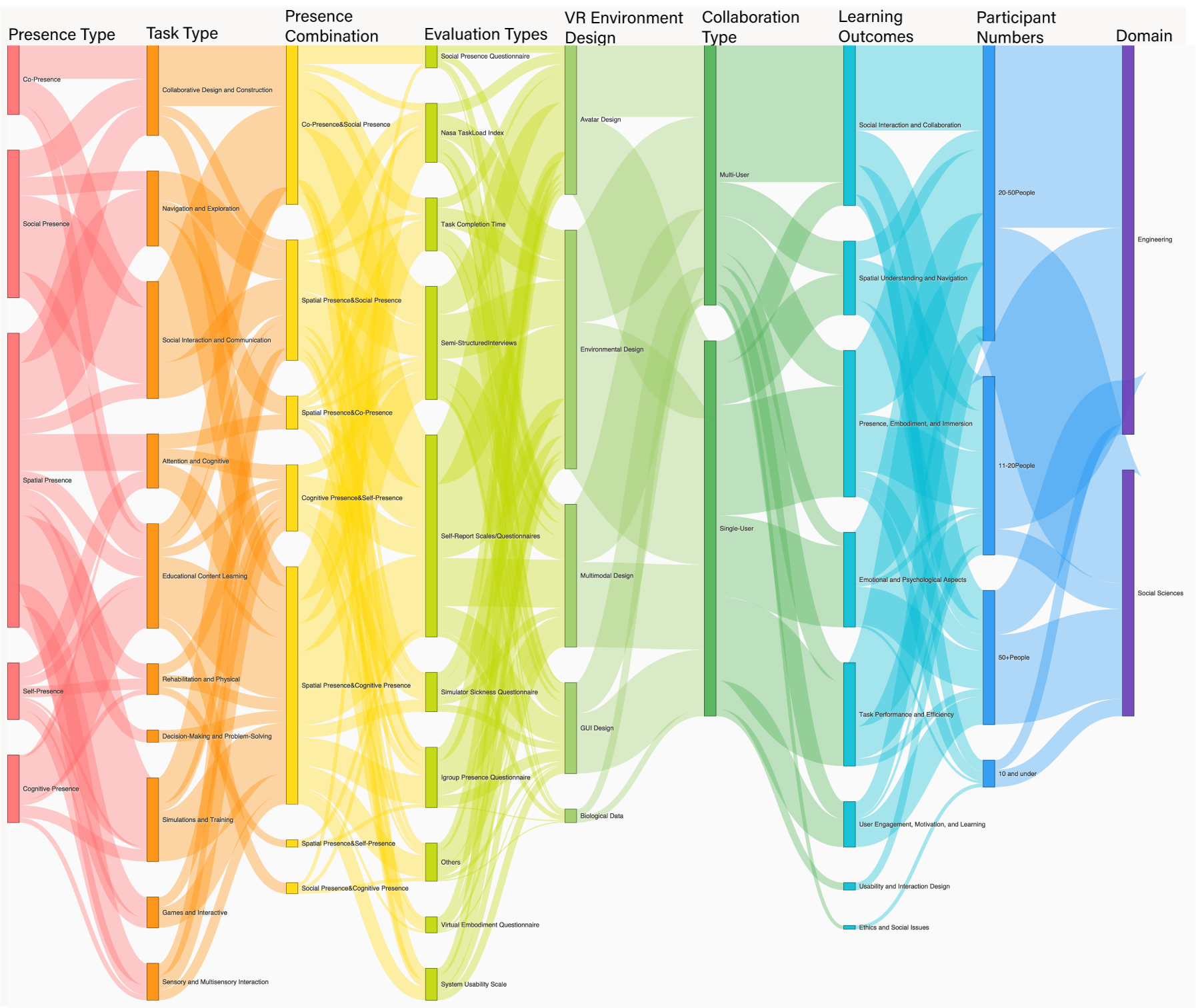}
    \caption{Sankey Diagram, a visualization showing the distribution of characteristics across all dimensions in VR research. Different stages (nodes) are color-coded to represent distinct dimensions: \includegraphics[height=0.25cm]{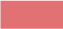} Presence Type, indicating various forms of presence experienced in VR (e.g., social, spatial, or cognitive presence); \includegraphics[height=0.25cm]{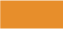} Task Type, categorizing specific tasks performed in VR environments; \includegraphics[height=0.25cm]{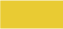} Presence Combination, highlighting interactions between multiple types of presence; \includegraphics[height=0.25cm]{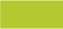} Evaluation Types, illustrating assessment methods such as questionnaires or task performance measures; \includegraphics[height=0.25cm]{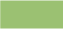} VR Environment Design, focusing on design elements like avatars or multimodal systems; \includegraphics[height=0.25cm]{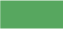} Collaboration Type, distinguishing between single-user and multi-user setups; \includegraphics[height=0.25cm]{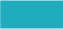} Learning Outcomes, emphasizing results such as social interaction or spatial understanding; \includegraphics[height=0.25cm]{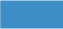} Participant Numbers, showcasing the size of participant groups; and \includegraphics[height=0.25cm]{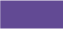} Domain, representing application areas such as engineering or social sciences. The width of the flows (links) is proportional to the total number of reviewed articles (N = 78), reflecting the prominence and interconnectedness of these research dimensions.}
    \label{XX}
\end{figure}

Based on the analytical results we have extracted from relevant literature, Figure \ref{XX} has provided an intuitive summary of the number of studies related to each dimension discussed in this article. The previous section has primarily elaborated on the factual content of the collected literature. Next, we explore these facts further, highlighting key research findings and insights. Therefore, this section thoroughly explores the common patterns and differences observed within these selected dimensions, focusing on how different types of presence in virtual reality (VR) environments have affected learning outcomes and task performance. Our discussion focuses on the following aspects:

\subsection{Moderating Effects of Design Factors on Presence and Learning Outcomes}
In VR environments, the complex relationship between presence and learning outcomes is significantly moderated by various design factors. These design elements not only shape the user's immersive experience but also profoundly influence the interaction between different types of presence and task performance as well as cognitive load. This section highlights key design components—multisensory feedback, virtual character design, self-presence and body perception design, and feedback mechanisms and interaction design—and examines how they moderate the relationship between presence and learning outcomes in different learning and collaborative contexts. By synthesizing the findings from various studies reviewed, we aim to reveal the pivotal role of thoughtful design in optimizing VR-based educational and collaborative experiences.

\subsubsection{Multi-Sensory Feedback Design}
Multisensory feedback plays a crucial role in enhancing the sense of presence within VR learning environments. The integration of multimodal stimuli such as visual, auditory, and tactile inputs can create a more immersive and interactive experience, thereby significantly enhancing both spatial presence and cognitive presence. Our research findings indicate that rich sensory input not only deepens users' cognitive engagement but also improves the accuracy and efficiency of task execution \cite{ lee2024may,schott2023vreal,peck2021evidence,aseeri2021influence,hu2021abio,thanyadit2023tutor}. For example, Schott et al. \cite{schott2023cardiogenesis4d} found in their VR simulation study on embryonic heart development that high-fidelity visual and tactile feedback significantly enhanced participants' spatial presence and facilitated better retention and understanding of information. Through tactile feedback, users are able to manipulate virtual objects more naturally, which not only increases the realism of the environment but also supports more efficient information processing and task completion.

Additionally, multisensory feedback plays a key role in managing cognitive load. Our research discovered that by providing diverse sensory cues, users can process information more efficiently, thereby reducing the consumption of cognitive resources \cite{sasikumar2024user,wong2023comparing,wei2023feeling,bellgardt2023virtual,michael2020race}. However, it is important to balance sensory inputs during the design process to prevent cognitive overload caused by excessive or incongruent sensory stimuli, which can undermine learning outcomes. Bueno-Vesga et al. \cite{bueno2021effects} pointed out that although multisensory feedback typically alleviates cognitive stress by providing clear and timely responses, overly complex or improperly integrated sensory elements may lead to user frustration and decreased task performance. Therefore, designers should carefully calibrate the intensity and coordination of each sensory input when developing multisensory feedback to ensure it aligns closely with learning objectives and task requirements, thereby avoiding interference from sensory information.

\subsubsection{Design of Avatar and Emotional Expression}
The design of avatars and their emotional expressions plays a critical role in shaping social presence and co-presence, especially in tasks that require high levels of collaboration and interactivity \cite{sasikumar2024user,unruh2023body,wong2023comparing,bhargava2023empirically,choudhary2023exploring}. Our research indicates that avatars with rich emotional behaviors and realistic interaction capabilities can significantly enhance users' social presence, thereby improving communication quality and collaboration efficiency \cite{peck2021evidence,aseeri2021influence,wang2023designing,espositi2024room,koulouris2020me,sykownik2022something,kimmel2023let,li2023we,shen2024legacysphere}. For example, Sasikumar et al. \cite{sasikumar2024user} demonstrated in their study how avatars with expressive facial and gesture animations enhanced users' co-presence, thereby accelerating task completion and strengthening social bonds among team members \cite{sasikumar2024user}. This finding is particularly important in multi-user training scenarios that require emotional engagement and empathy. Additionally, emotionally rich avatars can facilitate emotional contagion and the establishment of trust, which are crucial for maintaining team cohesion and increasing motivation. Choudhary et al. \cite{choudhary2023exploring} found that virtual humans conveying conflicting emotions significantly affected users' cognitive load by enhancing emotional resonance and trust. This suggests that carefully designed emotional expressions not only enhance the realism of interactions but also support users' mental well-being, thereby creating an environment conducive to learning and collaboration \cite{picard2000affective,pan2018and}. However, it is essential to ensure the appropriateness of emotional expressions in specific contexts during the design process to avoid inducing negative emotional states such as stress or anxiety, which could undermine learning outcomes \cite{choudhary2023exploring}.

\subsubsection{Self-Presence and Body-Perception Design}
Self-presence, which encompasses the user's bodily perception of their virtual avatar and the degree of alignment with it, is a crucial factor influencing motion control and overall realism in VR environments \cite{kim2023or,behm2013mirrored}. Our research indicates that the design of bodily perception mechanisms, such as full-body tracking and gesture recognition, directly determines the user's ability to interact naturally within the virtual space \cite{sasikumar2024user,yun2024exploring,yang2022hybridtrak,wang2023designing,kimmel2023let,espositi2024room,lin2020architect}. For example, Yang et al. \cite{yang2022hybridtrak} demonstrated how highly accurate full-body tracking systems significantly enhanced users' self-presence, thereby improving their motion performance and their sense of realism within the virtual environment. This high alignment between physical and virtual movements not only increases immersion but also facilitates skill acquisition and task proficiency by allowing users to repeatedly practice real-world actions in a safe and controlled virtual environment.

Our study also found that the design of self-presence mechanisms further regulates cognitive load by influencing how users process spatial and motion information \cite{bhargava2023empirically,voigt2021don,li2022synthesizing}. Enhanced self-presence can improve users' cognitive flexibility and adaptability, enabling them to engage more deeply in complex tasks and dynamic environments \cite{wijayanto2023comparing,michael2020race,koulouris2020me,jicol2021effects}. For instance, Wijayanto et al. \cite{wijayanto2023comparing} discovered that higher visual fidelity and accurate bodily perception systems significantly reduced users' cognitive load in spatial judgment tasks, thereby enhancing task performance and learning efficiency \cite{wijayanto2023comparing}. However, designers should be mindful of balancing the complexity of bodily perception systems to avoid issues such as motion sickness and psychological discomfort caused by overly complex systems, as highlighted by Pöhlmann et al. \cite{pohlmann2023you}. Therefore, optimizing bodily perception features to enhance self-presence while minimizing negative impacts on user comfort and cognitive load is an important consideration in the design process.

\subsubsection{Feedback Mechanisms and Interaction Design}
Effective feedback mechanisms and intuitive interaction designs have played a fundamental role in shaping users' sense of control and immersion within VR environments. Natural and responsive interaction methods, such as gesture control and voice commands, have significantly enhanced spatial presence and social presence, making users' actions more integrated into the virtual experience \cite{sasikumar2024user,mei2021cakevr,jicol2021effects,yun2024exploring,jetter2020vr}. For example, gesture control has enabled users to manipulate virtual objects with precision, increasing their sense of agency and task engagement. In the study by Mei et al. \cite{mei2021cakevr}, gesture coordination in a VR cake decorating task significantly enhanced users' sense of engagement and task performance. Feedback mechanisms have also played a critical role in reinforcing learning by providing immediate and clear responses to user behaviors, helping learners understand the consequences of their interactions, thereby promoting better decision-making and skill development. Jicol et al. \cite{jicol2021effects} demonstrated how a VR environment combining task-based questionnaires and real-time feedback mechanisms significantly improved users' emotional and cognitive presence, thereby enhancing overall learning outcomes. Additionally, integrating physiological feedback, such as heart rate and eye-tracking, has provided more profound insights into user states, allowing the learning experience to adapt to individual needs and cognitive loads. However, the design of feedback mechanisms should ensure that they are not overly intrusive or distracting, as excessive or poorly timed feedback can interrupt user focus and undermine the learning experience. Balancing the informativeness and subtlety of feedback has been crucial for maintaining an optimal flow state, enabling users to fully engage without being overwhelmed. As shown by Sasikumar et al. \cite{sasikumar2024user}, combining self-report scales with physiological data can provide a comprehensive assessment of user experience, thereby optimizing feedback mechanisms to effectively support presence and learning outcomes.

\subsection{The Multidimensional Influence of Presence in Engineering and Collaborative Tasks}

\subsubsection{The Direct and Indirect Effects of Presence}
Our research has found that in engineering and collaborative tasks, the impact of presence is both direct and indirect. The direct impact is primarily manifested in learners' sense of engagement and immersion in the task, both of which are core factors driving efficient learning and task execution \cite{schott2023cardiogenesis4d,wong2023comparing,michael2020race,dickinson2021diegetic,yun2024exploring,gu2022role,mal2023impact}. A high level of spatial presence has enabled learners to perceive and interact with the virtual environment more realistically, thereby improving their understanding and mastery of complex engineering concepts and operational processes \cite{wijayanto2023comparing,schott2023cardiogenesis4d,reichherzer2022supporting,sasikumar2024user,wei2023feeling}. For example, Wijayanto et al. \cite{wijayanto2023comparing} by comparing VR environments with different levels of visual fidelity, discovered that high spatial presence significantly enhanced users' performance in spatial navigation and object manipulation tasks.

However, the indirect impact has been facilitated by improving learners' emotional connections, social interactions, and team collaboration, thereby promoting better learning outcomes and task performance \cite{fidalgo2023magic}. Social presence and co-presence are particularly important in collaborative tasks, as they can strengthen trust and interaction among team members, thereby improving overall collaboration efficiency and innovation capability. Sasikumar et al. \cite{sasikumar2024user} have indicated in their study that enhanced social presence not only accelerated task completion but also reinforced social bonds among team members, thereby elevating the overall level of team collaboration \cite{sasikumar2024user,wong2023comparing,li2023we,jin2023collaborative,dickinson2021diegetic,wagner2021effect,cao2023dreamvr}. Furthermore, emotional connections, by increasing the emotional investment of learners, have further facilitated the internalization and application of knowledge, making the learning process more efficient and enduring \cite{unruh2023body}.

In summary, the role of presence in engineering and collaborative tasks is multi-layered and complex. The direct impact promotes learning and task execution by improving engagement and immersion; the indirect impact optimizes team collaboration and innovation capabilities by strengthening emotional connections and social interactions. Understanding these direct and indirect effects is of significant importance for designing effective VR learning and collaborative environments.

\subsubsection{Dynamic Adjustment and Adaptation of Presence}
Presence is not fixed but can be dynamically adjusted and adapted according to task demands and learner behavior. By designing adaptive virtual environments, various elements within the environment, such as the interaction modes of virtual characters and the complexity of tasks, can be adjusted based on learner feedback and actions, thereby enhancing the learner's immersion and engagement, and optimizing learning outcomes. A key method for dynamically adjusting presence is to optimize the environment and interactions based on real-time user feedback \cite{tian2023using,bhargava2023empirically}. For example, the system can monitor physiological indicators (such as heart rate and skin conductance) and behavioral data (such as operation speed and error rate) to dynamically adjust the intensity and type of sensory stimuli, maintaining an optimal level of presence and cognitive load. Sasikumar et al. \cite{sasikumar2024user} developed an adaptive VR learning system that automatically adjusts the difficulty of the task and the feedback frequency based on the user’s real-time performance, significantly improving the engagement of learners and the learning outcomes. In addition, dynamically adjusting task complexity is an important method for improving presence and learning outcomes. At different stages of learning or tasks, based on the progress and performance of the learner, the system can gradually increase or decrease task complexity to ensure that the learner is always in the optimal learning zone (the balance between challenge and ability) \cite{unruh2023body,tian2023using}. This dynamic adjustment not only maintains the learner's interest and motivation but also effectively prevents cognitive overload or learning fatigue. For example, Tian et al. \cite{tian2023using}, in their research on engineering design tasks, found that by dynamically adjusting task complexity and collaboration requirements, team collaboration efficiency and innovative results were significantly improved.

We maintain that dynamic adjustment and adaptation of presence are important strategies for achieving personalized and efficient learning. Through real-time monitoring and feedback, the virtual environment can be self-optimized according to the needs and behavior of the learner, improving immersion and participation, thus achieving better learning outcomes and task performance.

\subsubsection{The Universality and Specificity of Presence in Interdisciplinary Applications}

Our research indicates that the design and application of presence may exhibit different characteristics and requirements across various disciplines and tasks. Understanding the universality and specificity of presence helps designers tailor appropriate VR environments and interaction designs based on the specific needs of a given task or discipline, thereby maximizing learning and collaboration outcomes. In some fields, such as medicine and engineering, improving
spatial presence is particularly important. These disciplines often involve complex spatial cognition and precise operational skills, which require learners to accurately perceive and manipulate objects in the virtual environment \cite{schott2023cardiogenesis4d,bellgardt2023virtual}. Our research indicates that a high level of spatial presence significantly improves learners' precision and efficiency in these areas. For example, Schott et al. \cite{schott2023cardiogenesis4d} demonstrated in their medical training study that enhancing spatial presence significantly improved student performance and understanding in virtual anatomy tasks.

In other fields, such as language learning and artistic creation, social presence and emotional connection may play a more crucial role. These disciplines emphasize interactivity and emotional expression, requiring students to communicate and collaborate effectively with virtual characters or other learners in the virtual environment \cite{wei2023feeling,wei2024hearing,fidalgo2023magic,lee2024may,gu2022role}. By enhancing social presence, learners can naturally engage in interactions, improving their language skills and artistic creativity. For example, Gu et al. \cite{gu2022role} found in their language learning study that enhanced social presence significantly increased learners' speaking abilities and communication confidence.

Furthermore, the design of presence in interdisciplinary applications should consider the specific needs of the tasks. For example, in engineering design tasks, virtual environments need to provide high precision tools and material simulations to support complex design and modeling processes \cite{bellgardt2023virtual,tian2023using,sasikumar2024user,wijayanto2023comparing}, while in artistic creation tasks, the virtual environment should offer rich creative tools and flexible expressive means to inspire learners' creativity and expressive abilities \cite{wei2023feeling,wei2024hearing,wei2024multi}. Therefore, designers in interdisciplinary applications should flexibly adjust the focus of presence design according to the specific demands of the task, ensuring that the virtual environment effectively supports the learning and collaboration goals of different disciplines.

We suggest that the universality and specificity of presence in interdisciplinary applications reflect its diverse needs across different fields. By understanding and applying the various dimensions of presence, designers can tailor VR environments and interaction designs suited to the specific characteristics of each discipline and task, thereby enhancing learning and collaboration outcomes.

\subsection{Practical Application and Design Suggestions}

Based on the discussion above, we propose the following practical applications and design recommendations to optimize the design of VR learning environments and enhance learning outcomes:

\subsubsection{Design a Multi-Sensory Feedback System}
In designing multi-sensory feedback, the primary principle is to ensure that different sensory channels (visual, auditory, tactile, haptic feedback, etc.) are matched with specific educational objectives or operational tasks, thereby providing necessary and appropriate assistance at the sensory level. For example, for engineering or medical simulations that require strengthening spatial understanding, emphasis can be placed on enhancing visual accuracy and 3D tactile feedback; for tasks that need to reinforce emotional engagement or language expression, more attention can be paid to auditory cues and details such as micro-expressions and intonation \cite{lee2024may,schott2023vreal}.

To prevent the overlapping of multi-sensory information from causing cognitive overload, it is recommended to adopt a "layered integration" strategy in task flow design: first, provide high fidelity or strong feedback intensity in key stages to ensure that core information is prioritized; then, adopt weak feedback or delayed feedback mechanisms in secondary stages to reduce the learner's sensory load \cite{bueno2021effects,sasikumar2024user}. For example, in surgical simulations, stronger haptic feedback or voice prompts can be added during critical actions such as cutting and suturing, while maintaining low-intensity tactile and visual prompts during basic positioning or navigation processes.

To further enhance the effectiveness of multi-sensory feedback, physiological or behavioral monitoring methods (such as heart rate, motion rate, error rate, etc.) can be integrated to dynamically adjust the feedback intensity and presentation forms based on the learner's real-time state \cite{michael2020race,wong2023comparing}. For instance, when the system detects a significant increase in the learner's error rate in a particular stage, it can automatically enhance tactile or visual prompts to help them refocus; when the system recognizes that the learner has already mastered the relevant skills, it can appropriately reduce the feedback intensity, encouraging them to shift more attention to exploring new knowledge.

\subsubsection{Enhance Avatar Design}
Avatar emotional expression and cultural adaptation directly influence Social Presence and the quality of team collaboration \cite{choudhary2023exploring,li2023we}. Therefore, character design can incorporate facial expression recognition and big data analysis to provide personalized facial expressions, behaviors, and attire elements tailored to learners from different contexts and cultural backgrounds, achieving more inclusive and immersive social interactions. For specific academic or professional needs (such as nursing or language learning), rich character scripts can also be pre-designed, including common greetings and emotional feedback patterns, to meet the nuanced requirements of various tasks or scenarios.

In addition to traditional fixed animations or preset actions, real-time emotion computation and intent recognition technologies (such as voice semantic analysis and body posture capture) can be utilized to dynamically drive the emotional expressions of virtual characters. For example, in multi-user collaborations, if a tense team atmosphere or poor communication is detected, the Facilitator Avatar can alleviate the situation through a gentle tone and friendly expressions, and provide specific collaboration suggestions at the system level \cite{sasikumar2024user,bhargava2023empirically}. This dynamic and multi-dimensional emotional interaction will further enhance empathy and trust, promoting group cohesion and communication efficiency.

While striving for high-fidelity appearance and motion capture, it is also essential to carefully consider the potential disruptions caused by the Uncanny Valley effect \cite{kimmel2023let,yun2024exploring}. For learners, overly realistic character images with slight discrepancies in details or expressions can evoke strong discomfort or psychological burdens. It is recommended to conduct user testing across different tasks and user groups to determine the appropriate level of character realism and acceptable expression details, thereby finding a balance between ``realism'' and ``acceptability.''

\subsubsection{Optimize Collaborative Task Design}

To highlight the social presence and efficiency of collaboration in cooperative tasks, clear and complementary roles should be assigned to each participant based on learning objectives and the expertise of the members \cite{jin2023collaborative,wagner2021effect}. Additionally, task design should progress from simple to complex in a layered manner, breaking down complex tasks into several assessable phased objectives. This encourages team members to take turns assuming roles such as leader, recorder, executor, or supervisor, thereby forming an efficient task collaboration and problem-solving mechanism \cite{tian2023using,li2023we}.

Intuitive collaboration tools and interface elements, such as multi-user shared whiteboards, virtual rulers, and real-time annotation systems, should be used to assist teams in obtaining more direct feedback during interactions and decision-making processes \cite{cao2023dreamvr,unruh2023body}. Visualization components such as progress bars or task dashboards can help teams clearly understand current work progress, resource consumption, and member contributions, enabling timely strategy adjustments or task allocations. Furthermore, voice- or gesture-based quick positioning and marking features can enhance the shared understanding of collaboration goals, improving the efficiency of multiperson synchronized operations and conversations \cite{dickinson2021diegetic}.

For projects that require the participation of cross-regional or interdisciplinary teams, the customization and simulation advantages of VR environments can be fully utilized by integrating and visualizing key elements, processes, and tools from different disciplines or industries within the same virtual space \cite{gu2022role,wang2023designing}. By seamlessly switching scenes or interaction modes, users can experience multi-scene and multi-task collaborations in a single learning or training session, fostering more innovative interdisciplinary discussions and problem-solving approaches.

\subsubsection{Manage Cognitive Load}
In VR teaching or training, information presentation often exceeds the learner's immediate processing capacity. Therefore, during the content planning stage, it is necessary to prioritize knowledge points or skills, determining which information should be conveyed at critical moments and which can be temporarily hidden or simplified \cite{bueno2021effects,wijayanto2023comparing}. In actual interactions, the system can trigger the appearance of additional information based on conditions such as task completion status and time dimensions. For example, detailed explanations of principles or extended knowledge points can pop up after the user completes key steps, ensuring that learners receive the appropriate information at the right time.

By monitoring physiological or behavioral indicators (such as heart rate, operation speed, and error types), the system can dynamically adjust task difficulty or feedback intensity, keeping learners within a reasonable ``challenge-skill matching zone'' \cite{tian2023using,sasikumar2024user}. For example, when the system detects that a learner has high proficiency in a particular operation, it can automatically increase task complexity or shorten the operation time limit. Conversely, if the learner makes obvious errors, the system can reduce difficulty or provide immediate step-by-step instructions, thereby reducing frustration and avoiding cognitive collapse.

When assessing the impact of VR systems on cognitive load, evaluations should not be limited to self-assessment questionnaires or single behavioral indicators. A comprehensive approach should consider multiple dimensions of the data, including physiological data (such as EEG and EDA), interaction behavior logs (such as click trajectories and gaze tracking) and self-assessment scales, to form a complete cognitive load profile \cite{pohlmann2023you,sasikumar2024user}. Based on these diversified data, the system can achieve more precise dynamic interventions and optimizations, providing differentiated incentive strategies and difficulty adjustment schemes for different types of learners, effectively managing cognitive load at both the general and individual levels.

\section{CHALLENGES AND FUTURE DIRECTIONS}\label{section5}
Despite substantial progress in using VR to enhance presence and facilitate various tasks, several formidable challenges hinder the optimal deployment and scalability of VR technologies in educational and collaborative settings. This section highlightsthe primary technical, ethical, and practical impediments identified in the current literature and outlines prospective avenues for future research to address these issues comprehensively.

\subsection{Technical bottlenecks and scalability}

Some technologies limit the growth and widespread adoption of virtual reality and affect both the performance and scalability of virtual reality applications.

\subsubsection{Hardware Limitations} High-fidelity VR experiences require environments that are immersive and responsive, supported by advanced hardware. Although current VR headsets are continually improving, they still suffer from issues such as heavy weight, poor ergonomics, heat management, and limited battery life, leading to discomfort and fatigue during prolonged use \cite{biener2022quantifying}. Additionally, the high cost of VR systems poses barriers, especially for underdeveloped regions and budget-constrained non-profit educational institutions. Future research should prioritize the development of lightweight, ergonomically optimized, and cost-effective VR hardware to enhance accessibility and user comfort.

\subsubsection{Delayed and real-time processing} Achieving minimal latency is crucial for VR technology. Minimizing latency is essential to maintain immersion and prevent motion sickness. However, when dealing with real-time rendering of complex virtual environments or multi-user learning settings, maintaining low latency remains a significant challenge. Advances in edge computing and the integration of 5G networks offer promising solutions to reduce latency, but these technologies are still in the early stages of deployment. More improvements are needed to accommodate different network environments and meet the strict requirements of VR applications.

\subsubsection{Tracking Accuracy and Robustness} The accuracy of immersion and natural interaction in VR environments significantly affects the user experience of movements and interactions. Current tracking systems rely on a combination of external sensors and internal algorithms, which makes them susceptible to errors caused by environmental factors such as lighting conditions and physical obstacles \cite{niehorster2017accuracy,yang2022hybridtrak}. By integrating machine learning algorithms and sensor fusion technologies to improve the accuracy and robustness of tracking systems, the user experience and fidelity of interaction can be substantially improved \cite{wu2023virtual}.

\subsubsection{Content Scalability and Adaptability} Creating scalable and adaptable virtual reality content to meet diverse educational and collaborative needs is a long-standing challenge. Developing modular and reusable content frameworks can facilitate the rapid deployment and customization of VR experiences across different disciplines \cite{ponder2003vhd++}. Furthermore, using computer-generated content and AI-based adjustments can create dynamic and responsive VR environments that meet the personal needs and preferences of each user \cite{chamola2024comprehensive}.

\subsubsection{Interoperability and Standardization} The lack of standardized protocols and interoperability among different VR platforms and tools hinders the seamless integration and collaborative potential of VR applications. Our analysis of 78 articles indicates that VR applications are primarily developed on single platforms, with few cross-platform initiatives. Therefore, establishing universal standards for data formats, communication protocols, and interaction paradigms is crucial to building a cohesive VR ecosystem that supports interoperability, content sharing, and collaborative development.

\subsection{Ethical and social implications}
\subsubsection{Mental Health and Well-being} Prolonged exposure to VR environments can have adverse effects on mental health, including increased anxiety, disorientation, and motion sickness \cite{souchet2023narrative}. Furthermore, users may become overly dependent on virtual interactions, neglecting participation in the real world, making it necessary to carefully consider the potential for VR addiction \cite{han2022virtual}. Future research should investigate the long-term psychological impacts of using virtual reality technology and develop guidelines and interventions to mitigate the negative effects.

\subsubsection{Ethical Content and Representation} When designing virtual reality environments, it is essential to remain sensitive to cultural, social, and individual differences to prevent the perpetuation of biases and stereotypes \cite{taylor2020interracial,klein2024ethnic}. Ensuring diverse representations and fostering an inclusive environment in virtual reality can enhance user experience and promote social cohesion. For example, incorporating diverse character models, varied cultural content, and inclusive interaction paradigms can enhance the user experience and promote social cohesion \cite{schwind2017these,combe2024exploring,shadiev2021cross,creed2024inclusive}.

\subsubsection{Impact on Social Interactions} Balancing the benefits of virtual interactions with maintaining real-world social connections is crucial to promoting healthy social dynamics. The immersive nature of VR can alter the dynamics of social interactions, potentially leading to both positive and negative outcomes \cite{oh2023social}. Although virtual reality can facilitate meaningful connections and collaborative efforts, it can also result in social isolation and reduced face-to-face interactions \cite{della2022interpersonal}. Developing comprehensive ethical frameworks, promoting inclusive design practices, and fostering ongoing dialogues about the responsible use of VR technology are key steps in mitigating potential harms and ensuring equitable benefits.

\subsection{Unleashing the potential of VR in distance learning}

VR transcends the traditional boundaries of online education, positioning itself at the forefront of remote learning innovation. Unlike conventional platforms that rely on two-dimensional interactions, VR immerses learners in three-dimensional environments, fostering a sense of presence that closely mirrors face-to-face experiences. This immersion is not just a technological novelty, but a transformative tool that significantly enhances cognitive engagement and memory retention. Furthermore, VR promotes experiential learning, allowing learners to interact directly with the material in ways that static images or videos cannot replicate. This hands-on approach deepens understanding and cultivates critical thinking and problem solving skills essential for real-world applications. The ability to simulate real-life scenarios in VR also prepares learners to tackle practical challenges without the associated risks or costs, making education more accessible and equitable. Importantly, VR improves social connections and collaborative learning. Virtual classrooms equipped with avatars and spatial audio create environments where learners can interact naturally, share ideas, and collaborate on projects as if they were physically together. This social presence mitigates the isolation often felt in remote learning settings, fostering a more engaging and supportive educational experience. To fully realize these advantages, it is crucial to address barriers such as accessibility, affordability, and the development of VR content aligned with high-quality curriculum. In addition, educators should be trained to effectively integrate VR into their teaching methodologies, ensuring that technology enhances rather than distracts from learning objectives. As VR technology continues to advance, its integration into remote education promises not only to improve educational outcomes, but also to democratize access to high-quality education globally.

\subsection{Vertical research and practical deployment}
To fully unlock the potential of VR, research should focus on specific industries and applications. Vertical research addresses the unique needs and challenges of fields such as healthcare, engineering, and the arts, allowing the creation of customized VR solutions. Successful deployment of VR requires collaboration between academia, industry and developers. This partnership ensures that VR applications are not only theoretically sound, but also practical and scalable. Establishing industry-specific standards and best practices can facilitate the seamless integration of VR into existing workflows and educational programs, enhancing adoption rates and user satisfaction. User-centric design is essential for creating intuitive, accessible, and adaptable VR solutions. Incorporating continuous feedback from end-users helps refine VR experiences, making them more effective and engaging. In addition, addressing scalability issues, such as improving hardware accessibility and ensuring software interoperability, is crucial for widespread use.


\section{CONCLUSION}\label{section6}

This systematic review analyzes 78 articles to examine how various types of presence in VR—spatial, social, co-presence, cognitive, and self-presence—affect remote collaboration and educational tasks. The findings imply that spatial presence enhances immersion and interaction realism, making users feel genuinely part of the virtual environment. Social and co-presence significantly boost communication and team cohesion, fostering stronger connections and collaboration among users. Cognitive presence deepens engagement and problem-solving abilities, facilitating effective task immersion and teamwork. Also, self-presence reinforces user identity and emotional bonds, making virtual environments more personalized and meaningful. The research underscores the necessity of comprehensive VR design that incorporates realistic environments, sensory inputs, avatars, and adaptive feedback to optimize learning and collaboration outcomes. However, it also highlights challenges such as technical limitations—including hardware constraints and the need for more precise tracking systems—and ethical issues like privacy concerns and the psychological effects of prolonged immersion. To address these challenges, the study advocates for continual advancements in VR technologies and the establishment of ethical guidelines to ensure user safety and equity. Additionally, it recommends future research to explore the interplay of different presence types and their combined impact on complex tasks within diverse educational and collaborative settings. Ultimately, VR has the potential to significantly enhance education and teamwork by providing more engaging, effective, and inclusive digital environments. As VR technology advances, its applications across various domains—from remote work to immersive education—are likely to influence how people interact, learn, and collaborate in the digital age.

\bibliographystyle{ACM-Reference-Format}
\bibliography{Ref}

\clearpage
\appendix

\section{A DATA EXTRACTION LIST FOR INCLUDED PUBLICATION}\label{A}

\begin{table}[htp]
\centering
\caption{Presence Type and Task Type (DE4 and DE5)}
\scalebox{0.67}{
\begin{tabular}{cl}
\hline
Category                                                                                         & Citations \\ \hline
\textbf{\begin{tabular}[c]{@{}c@{}}Presence Type\\ (DE4) \end{tabular}}                                                                       &          \\
\multirow{2}{*}{Spatial Presence}                                                                &\cite{yun2024exploring}\cite{wijayanto2023comparing}\cite{bhargava2023empirically}\cite{reichherzer2022supporting}\cite{wagner2021effect}\cite{schott2021vr}\cite{dickinson2021diegetic}\cite{voigt2021don}\cite{peck2021evidence}\cite{li2020analysing}\cite{hsu2020design}\cite{wei2023feeling}\cite{wei2024hearing}\cite{hu2021abio}\cite{kim2022multi}\cite{lin2020architect}\cite{jetter2020vr}\cite{yang2022hybridtrak}\cite{thoravi2022dreamstream}\cite{tseng2022dark}\cite{pohlmann2023you}\cite{cao2023dreamvr}\cite{luo2024exploring}            \\
                                                                                                 & \cite{gomi2024inflatablebots}\cite{dufresne2024touching}\cite{kargut2024effects}\cite{gottsacker2021diegetic}\cite{schott2023vreal}\cite{bartl2022effects}\cite{wang2022realitylens}\cite{10.1145/3641825.3687735}\cite{10.1145/3641825.3687735}\cite{10.1145/3492802}         \\
Social Presence                                                                              & \cite{schott2024excuse}\cite{lee2024may}\cite{wong2023comparing}\cite{choudhary2023exploring}\cite{bozkir2021exploiting}\cite{aseeri2021influence}\cite{wang2023designing}\cite{jin2024headsetoff}\cite{espositi2024room}\cite{mei2021cakevr}\cite{petersen2021pedagogical}\cite{sykownik2022something}\cite{kimmel2023let}\cite{o2023re}\cite{jin2023collaborative}\cite{li2023we}\cite{shen2024legacysphere}\cite{wang2024socially}\cite{chen2024d}\cite{yoon2020evaluating}\cite{10.1145/3686910}\cite{moss2024going}     \\
                                                          
Co-presence                                                                                      & \cite{sasikumar2024user}\cite{fidalgo2023magic}\cite{tian2023using}\cite{volonte2021effects}\cite{thanyadit2023tutor}\cite{wei2023bridging}\cite{yakura2020enhancing}\cite{abramczuk2023meet}
                                                                   \\
Self-presence                                                                                    & \cite{unruh2023body}\cite{gu2022role}\cite{mal2023impact}\cite{wolf2021embodiment}\cite{michael2020race}\cite{bahng2020reflexive}\cite{koulouris2020me}    \\
Cognitive Presence                                                                               & \cite{schott2023cardiogenesis4d}\cite{bellgardt2023virtual}\cite{li2022synthesizing}\cite{bueno2021effects}\cite{hamzeheinejad2021impact}\cite{shigyo2024vr}\cite{jicol2021effects}\cite{jicol2023realism} \\      

\textbf{\begin{tabular}[c]{@{}c@{}}Task Type \\ (DE5)\end{tabular}}  &          \\
\begin{tabular}[c]{@{}c@{}}1.Collaborative design \\ and construction tasks\end{tabular}           & \cite{sasikumar2024user}\cite{wong2023comparing}\cite{tian2023using}\cite{hsu2020design}\cite{lin2020architect}\cite{mei2021cakevr}\cite{10.1145/3686910}\cite{abramczuk2023meet}         \\

\begin{tabular}[c]{@{}c@{}}2.Navigation and \\ exploration tasks\end{tabular}                      & \cite{schott2024excuse}\cite{yun2024exploring}\cite{wagner2021effect}\cite{luo2024exploring}\cite{10.1145/3641825.3687735}     \\

\begin{tabular}[c]{@{}c@{}}3.Simulations and \\ training tasks\end{tabular}                        & \cite{li2022synthesizing}\cite{dickinson2021diegetic}\cite{peck2021evidence}\cite{li2020analysing}\cite{wei2024hearing}\cite{jetter2020vr}\cite{koulouris2020me}\cite{jicol2021effects}\cite{dufresne2024touching}\cite{schott2023vreal}          \\


\begin{tabular}[c]{@{}c@{}}4.Social interaction and \\ communication tasks\end{tabular}            & \cite{lee2024may}\cite{choudhary2023exploring}\cite{fidalgo2023magic}\cite{volonte2021effects}\cite{aseeri2021influence}\cite{wang2023designing}\cite{jin2024headsetoff}\cite{espositi2024room}\cite{sykownik2022something}\cite{thoravi2022dreamstream}\cite{kimmel2023let}\cite{o2023re}\cite{wei2023bridging}\cite{li2023we}\cite{shen2024legacysphere}\cite{wang2024socially}\cite{chen2024d}\cite{yakura2020enhancing}\cite{yoon2020evaluating}\cite{wang2022realitylens}\cite{moss2024going}            \\


\begin{tabular}[c]{@{}c@{}}5.Games and \\ interactive tasks\end{tabular}                           & \cite{voigt2021don}\cite{michael2020race}\cite{10.1145/3641825.3687753}        \\


\begin{tabular}[c]{@{}c@{}}6.Educational content \\ exploration \\ and learning tasks\end{tabular} & \cite{schott2023cardiogenesis4d}\cite{bellgardt2023virtual}\cite{gu2022role}\cite{schott2021vr}\cite{bozkir2021exploiting}\cite{bueno2021effects}\cite{wei2023feeling}\cite{bahng2020reflexive}\cite{petersen2021pedagogical}\cite{thanyadit2023tutor}\cite{cao2023dreamvr}\cite{jin2023collaborative}             \\


\begin{tabular}[c]{@{}c@{}}7.Sensory and multisensory \\ interaction tasks\end{tabular}            & \cite{wolf2021embodiment}\cite{hu2021abio}\cite{kim2022multi}\cite{shigyo2024vr}\cite{yang2022hybridtrak}\cite{tseng2022dark}\cite{gomi2024inflatablebots}\cite{10.1145/3492802}         \\

\begin{tabular}[c]{@{}c@{}}8.Decision-making and \\ problem-solving tasks\end{tabular}             & \cite{reichherzer2022supporting}\cite{kargut2024effects}         \\

\begin{tabular}[c]{@{}c@{}}9.Rehabilitation and \\ physical tasks\end{tabular}                     & \cite{bhargava2023empirically}\cite{mal2023impact}\cite{hamzeheinejad2021impact}\cite{bartl2022effects}         \\

\begin{tabular}[c]{@{}c@{}}10.Attention and \\ cognitive tasks\end{tabular}                         & \cite{unruh2023body}\cite{wijayanto2023comparing}\cite{pohlmann2023you}\cite{jicol2023realism}\cite{gottsacker2021diegetic}        \\ \hline
\end{tabular}
}
\label{appendix-a}
\end{table}

\begin{table}[htp]
\centering
\caption{Evaluation Types and VR Environment Design (DE7 and DE8)}
\scalebox{0.67}{
\begin{tabular}{cl}
\hline
Category                                                                                         & Citations \\ \hline
\textbf{\begin{tabular}[c]{@{}c@{}}Evaluation Types \\ (DE7) \end{tabular}}                                                                       &          \\
Social Presence Questionnaire                                                                &\cite{sasikumar2024user}\cite{fidalgo2023magic}\cite{tian2023using}\cite{volonte2021effects}\cite{petersen2021pedagogical}\cite{sykownik2022something}\cite{abramczuk2023meet}         \\

Nasa TaskLoad Index                                                                             & \cite{sasikumar2024user}\cite{schott2024excuse}\cite{yun2024exploring}\cite{schott2023cardiogenesis4d}\cite{wijayanto2023comparing}\cite{tian2023using}\cite{li2022synthesizing}\cite{wagner2021effect}\cite{dickinson2021diegetic}\cite{bueno2021effects}\cite{li2020analysing}\cite{hsu2020design}\cite{lin2020architect}\cite{thoravi2022dreamstream}\cite{tseng2022dark}\cite{pohlmann2023you}\cite{dufresne2024touching}\cite{bartl2022effects}\cite{10.1145/3641825.3687753}     \\
                                                          
Task Completion Time                                                                                     & \cite{sasikumar2024user}\cite{schott2024excuse}\cite{yun2024exploring}\cite{lee2024may}\cite{wijayanto2023comparing}\cite{wong2023comparing}\cite{tian2023using}\cite{li2022synthesizing}\cite{dickinson2021diegetic}\cite{kim2022multi}\cite{lin2020architect}\cite{jetter2020vr}\cite{kargut2024effects}\cite{gottsacker2021diegetic}\cite{yoon2020evaluating}\cite{schott2023vreal}\cite{10.1145/3641825.3687735}
                                                                   \\
\multirow{2}{*}{Semi-Structured Interviews}                                                                                    & \cite{sasikumar2024user}\cite{schott2024excuse}\cite{schott2023cardiogenesis4d}\cite{schott2021vr}\cite{dickinson2021diegetic}\cite{li2020analysing}\cite{wei2023feeling}\cite{wei2024hearing}\cite{kim2022multi}\cite{hamzeheinejad2021impact}\cite{wang2023designing}\cite{jin2024headsetoff}\cite{shigyo2024vr}\cite{espositi2024room}\cite{michael2020race}\cite{bahng2020reflexive}\cite{lin2020architect}\cite{jetter2020vr}\cite{koulouris2020me}\cite{mei2021cakevr}\cite{thoravi2022dreamstream}\cite{tseng2022dark}\cite{thanyadit2023tutor}\cite{cao2023dreamvr}\cite{jin2023collaborative}\cite{wei2023bridging}\\ &\cite{li2023we}\cite{shen2024legacysphere}\cite{luo2024exploring}\cite{gomi2024inflatablebots}\cite{wang2024socially}\cite{dufresne2024touching}\cite{chen2024d}\cite{gottsacker2021diegetic}\cite{yoon2020evaluating}\cite{wang2022realitylens}\cite{10.1145/3686910}\cite{moss2024going}   \\

\multirow{3}{*}{Self-Report Scales/Questionnaires}                                                                               & \cite{schott2024excuse}\cite{yun2024exploring}\cite{lee2024may}\cite{unruh2023body}\cite{schott2023cardiogenesis4d}\cite{wijayanto2023comparing}\cite{wong2023comparing}\cite{bhargava2023empirically}\cite{choudhary2023exploring}\cite{bellgardt2023virtual}\cite{fidalgo2023magic}\cite{gu2022role}\cite{mal2023impact}\cite{reichherzer2022supporting}\cite{li2022synthesizing}\cite{wagner2021effect}\cite{schott2021vr}\cite{voigt2021don}\cite{volonte2021effects}\cite{peck2021evidence}\cite{bozkir2021exploiting}\cite{bueno2021effects}\cite{wolf2021embodiment}\cite{aseeri2021influence}\cite{li2020analysing}\cite{wei2023feeling}\\ &\cite{wei2024hearing}\cite{hu2021abio}\cite{kim2022multi}\cite{hamzeheinejad2021impact}\cite{wang2023designing}\cite{jin2024headsetoff}\cite{shigyo2024vr}\cite{espositi2024room}\cite{michael2020race}\cite{bahng2020reflexive}\cite{lin2020architect}\cite{jetter2020vr}\cite{koulouris2020me}\cite{mei2021cakevr}\cite{jicol2021effects}\cite{petersen2021pedagogical}\cite{sykownik2022something}\cite{yang2022hybridtrak}\cite{thoravi2022dreamstream}\cite{tseng2022dark}\cite{kimmel2023let}\cite{o2023re}\cite{jin2023collaborative}\cite{wei2023bridging}\cite{jicol2023realism}\cite{luo2024exploring}\\ &\cite{gomi2024inflatablebots}\cite{wang2024socially}\cite{chen2024d}\cite{kargut2024effects}\cite{gottsacker2021diegetic}\cite{yakura2020enhancing}\cite{yoon2020evaluating}\cite{schott2023vreal}\cite{bartl2022effects}\cite{wang2022realitylens}\cite{10.1145/3641825.3687735}\cite{10.1145/3641825.3687753}\cite{10.1145/3686910}\cite{10.1145/3492802}\cite{moss2024going}\cite{abramczuk2023meet} \\

Simulator Sickness Questionnaire                                                                              & \cite{schott2024excuse}\cite{yun2024exploring}\cite{unruh2023body}\cite{wong2023comparing}\cite{bellgardt2023virtual}\cite{mal2023impact}\cite{wagner2021effect}\cite{wolf2021embodiment}\cite{hsu2020design}\cite{pohlmann2023you}\cite{luo2024exploring}\cite{10.1145/3641825.3687735} \\      

Igroup Presence Questionnaire                                                                              & \cite{yun2024exploring}\cite{lee2024may}\cite{unruh2023body}\cite{schott2023cardiogenesis4d}\cite{wijayanto2023comparing}\cite{bhargava2023empirically}\cite{wagner2021effect}\cite{schott2021vr}\cite{dickinson2021diegetic}\cite{voigt2021don}\cite{bozkir2021exploiting}\cite{wolf2021embodiment}\cite{aseeri2021influence}\cite{pohlmann2023you}\cite{o2023re}\cite{dufresne2024touching}\cite{schott2023vreal}\cite{wang2022realitylens}\cite{10.1145/3641825.3687753}\cite{moss2024going} \\      

Virtual Embodiment Questionnaire                                                                              & \cite{unruh2023body}\cite{mal2023impact}\cite{dufresne2024touching}\cite{bartl2022effects} \\      

System Usability Scale                                                                            & \cite{schott2023cardiogenesis4d}\cite{bellgardt2023virtual}\cite{tian2023using}\cite{wagner2021effect}\cite{schott2021vr}\cite{hsu2020design}\cite{mei2021cakevr}\cite{thanyadit2023tutor}\cite{dufresne2024touching}\cite{10.1145/3641825.3687753} \\      

Others                                                                            & \cite{reichherzer2022supporting}\cite{voigt2021don}\cite{peck2021evidence}\cite{jicol2021effects}\cite{petersen2021pedagogical}\cite{yang2022hybridtrak}\cite{kargut2024effects}\cite{gottsacker2021diegetic}\cite{10.1145/3641825.3687735}\cite{hu2021abio} \\

\textbf{\begin{tabular}[c]{@{}c@{}}VR Environment Design  \\ (DE8)\end{tabular}}  &          \\

\begin{tabular}[c]{@{}c@{}}\multirow{2}{*}{Avatar Design} \end{tabular}           & \cite{sasikumar2024user}\cite{schott2024excuse}\cite{yun2024exploring}\cite{lee2024may}\cite{unruh2023body}\cite{schott2023cardiogenesis4d}\cite{wong2023comparing}\cite{bhargava2023empirically}\cite{choudhary2023exploring}\cite{bellgardt2023virtual}\cite{fidalgo2023magic}\cite{gu2022role}\cite{mal2023impact}\cite{tian2023using}\cite{voigt2021don}\cite{volonte2021effects}\cite{peck2021evidence}\cite{bozkir2021exploiting}\cite{bueno2021effects}\cite{wolf2021embodiment}\cite{aseeri2021influence}\cite{wang2023designing}\cite{jin2024headsetoff}\cite{shigyo2024vr}\cite{espositi2024room}\cite{michael2020race}\\ &\cite{koulouris2020me}\cite{mei2021cakevr}\cite{petersen2021pedagogical}\cite{sykownik2022something}\cite{thoravi2022dreamstream}\cite{kimmel2023let}\cite{thanyadit2023tutor}\cite{cao2023dreamvr}\cite{jin2023collaborative}\cite{wei2023bridging}\cite{shen2024legacysphere}\cite{wang2024socially}\cite{dufresne2024touching}\cite{chen2024d}\cite{gottsacker2021diegetic}\cite{bartl2022effects}\cite{wang2022realitylens}\cite{10.1145/3686910}\cite{abramczuk2023meet}                \\

\begin{tabular}[c]{@{}c@{}}\multirow{3}{*}{Environmental Design} \end{tabular}                      & \cite{sasikumar2024user}\cite{schott2024excuse}\cite{yun2024exploring}\cite{lee2024may}\cite{unruh2023body}\cite{schott2023cardiogenesis4d}\cite{wijayanto2023comparing}\cite{wong2023comparing}\cite{bhargava2023empirically}\cite{choudhary2023exploring}\cite{bellgardt2023virtual}\cite{fidalgo2023magic}\cite{gu2022role}\cite{mal2023impact}\cite{tian2023using}\cite{reichherzer2022supporting}\cite{li2022synthesizing}\cite{wagner2021effect}\cite{schott2021vr}\cite{dickinson2021diegetic}\cite{voigt2021don}\cite{volonte2021effects}\cite{peck2021evidence}\cite{bozkir2021exploiting}\cite{bueno2021effects}\\ &\cite{wolf2021embodiment}\cite{li2020analysing}\cite{hsu2020design}\cite{wei2023feeling}\cite{wei2024hearing}\cite{hu2021abio}\cite{kim2022multi}\cite{hamzeheinejad2021impact}\cite{wang2023designing}\cite{jin2024headsetoff}\cite{shigyo2024vr}\cite{espositi2024room}\cite{michael2020race}\cite{bahng2020reflexive}\cite{lin2020architect}\cite{jetter2020vr}\cite{mei2021cakevr}\cite{jicol2021effects}\cite{petersen2021pedagogical}\cite{thoravi2022dreamstream}\cite{tseng2022dark}\cite{pohlmann2023you}\cite{thanyadit2023tutor}\cite{cao2023dreamvr}\cite{jin2023collaborative}\\ &\cite{wei2023bridging}\cite{jicol2023realism}\cite{li2023we}\cite{shen2024legacysphere}\cite{luo2024exploring}\cite{gomi2024inflatablebots}\cite{wang2024socially}\cite{dufresne2024touching}\cite{chen2024d}\cite{kargut2024effects}\cite{gottsacker2021diegetic}\cite{schott2023vreal}\cite{bartl2022effects}\cite{wang2022realitylens}\cite{10.1145/3641825.3687735}\cite{10.1145/3641825.3687753}\cite{10.1145/3686910}\cite{10.1145/3492802}\cite{moss2024going}\cite{abramczuk2023meet}       \\

\begin{tabular}[c]{@{}c@{}} \multirow{2}{*}{Multimodal Design} \end{tabular}                        & \cite{sasikumar2024user}\cite{schott2024excuse}\cite{yun2024exploring}\cite{lee2024may}\cite{wijayanto2023comparing}\cite{wong2023comparing}\cite{bhargava2023empirically}\cite{choudhary2023exploring}\cite{bellgardt2023virtual}\cite{fidalgo2023magic}\cite{gu2022role}\cite{mal2023impact}\cite{tian2023using}\cite{reichherzer2022supporting}\cite{li2022synthesizing}\cite{wagner2021effect}\cite{voigt2021don}\cite{volonte2021effects}\cite{peck2021evidence}\cite{bozkir2021exploiting}\cite{bueno2021effects}\\ &\cite{li2020analysing}\cite{wei2024hearing}\cite{hu2021abio}\cite{wang2023designing}\cite{jin2024headsetoff}\cite{espositi2024room}\cite{bahng2020reflexive}\cite{lin2020architect}\cite{jetter2020vr}\cite{jicol2021effects}\cite{petersen2021pedagogical}\cite{yang2022hybridtrak}\cite{pohlmann2023you}\cite{o2023re}\cite{thanyadit2023tutor}\cite{jin2023collaborative}\cite{wei2023bridging}\cite{jicol2023realism}\cite{li2023we}\cite{luo2024exploring}\cite{gomi2024inflatablebots}\cite{dufresne2024touching}\cite{gottsacker2021diegetic}\cite{schott2023vreal}            \\

\begin{tabular}[c]{@{}c@{}}GUI Design\end{tabular}            & \cite{unruh2023body}\cite{schott2023cardiogenesis4d}\cite{wijayanto2023comparing}\cite{bellgardt2023virtual}\cite{wagner2021effect}\cite{schott2021vr}\cite{dickinson2021diegetic}\cite{volonte2021effects}\cite{hsu2020design}\cite{wei2023feeling}\cite{wei2024hearing}\cite{kim2022multi}\cite{hamzeheinejad2021impact}\cite{michael2020race}\cite{lin2020architect}\cite{mei2021cakevr}\cite{thanyadit2023tutor}\cite{o2023re}\cite{luo2024exploring}\cite{kargut2024effects}\cite{yakura2020enhancing}\cite{yoon2020evaluating}\cite{schott2023vreal}              \\

\begin{tabular}[c]{@{}c@{}}Biometric Data\end{tabular}                           & \cite{voigt2021don}\cite{jin2024headsetoff}\cite{michael2020race}\cite{lin2020architect}\cite{peck2021evidence}\cite{wang2023designing}        \\

\hline
\end{tabular}
}
\label{appendix-a}
\end{table}

\begin{table}[htp]
\centering
\caption{Presence Combination and Learning Outcomes (DE6 and DE10)}
\scalebox{0.67}{
\begin{tabular}{cl}
\hline
Category                                                                                         & Citations \\ \hline
\textbf{\begin{tabular}[c]{@{}c@{}}Presence Combination \\ (DE6) \end{tabular}}                                                                       &          \\
\multirow{2}{*}{Co-Presence \& Social Presence}                                                              &\cite{sasikumar2024user}\cite{lee2024may}\cite{wong2023comparing}\cite{choudhary2023exploring}\cite{fidalgo2023magic}\cite{tian2023using}\cite{volonte2021effects}\cite{aseeri2021influence}\cite{wang2023designing}\cite{jin2024headsetoff}\cite{mei2021cakevr}\cite{sykownik2022something}\cite{kimmel2023let}\cite{thanyadit2023tutor}\cite{jin2023collaborative}\cite{wei2023bridging}\cite{li2023we}\cite{shen2024legacysphere}\cite{chen2024d}\\

&\cite{yakura2020enhancing}\cite{yoon2020evaluating}\cite{10.1145/3686910}\cite{abramczuk2023meet}         \\

Spatial Presence \& Social Presence                                                                               &\cite{schott2024excuse}\cite{bozkir2021exploiting}\cite{li2020analysing}\cite{hsu2020design}\cite{kim2022multi}\cite{espositi2024room}\cite{thoravi2022dreamstream}\cite{o2023re}\cite{luo2024exploring}\cite{gomi2024inflatablebots}\cite{wang2024socially}\cite{dufresne2024touching}\cite{wang2022realitylens}\cite{moss2024going}     \\
                                                          
Spatial Presence \& Co-Presence                                                                                       &\cite{yun2024exploring}\cite{gottsacker2021diegetic} 
                                                                   \\
Cognitive Presence \& Self-presence                                                                                     &\cite{unruh2023body}\cite{gu2022role}\cite{mal2023impact}\cite{wolf2021embodiment}\cite{shigyo2024vr}\cite{michael2020race}\cite{bahng2020reflexive}\cite{koulouris2020me}\cite{jicol2021effects}   \\

\multirow{2}{*}{Spatial Presence \& Cognitive Presence}                                                                                &\cite{hu2021abio}\cite{schott2023cardiogenesis4d}\cite{wijayanto2023comparing}\cite{bellgardt2023virtual}\cite{reichherzer2022supporting}\cite{li2022synthesizing}\cite{wagner2021effect}\cite{schott2021vr}\cite{dickinson2021diegetic}\cite{voigt2021don}\cite{peck2021evidence}\cite{bueno2021effects}\cite{wei2023feeling}\cite{hamzeheinejad2021impact}\cite{wei2024hearing}\cite{lin2020architect}\cite{jetter2020vr}\cite{yang2022hybridtrak}\cite{tseng2022dark}\\

&\cite{pohlmann2023you}\cite{cao2023dreamvr}\cite{jicol2023realism}\cite{kargut2024effects}\cite{schott2023vreal}\cite{bartl2022effects}\cite{10.1145/3641825.3687735}\cite{10.1145/3641825.3687753}\cite{10.1145/3492802}  \\  

Spatial Presence \& Self-Presence                                                                                &\cite{bhargava2023empirically}  \\

Social Presence \& Cognitive Presence                                                                                &\cite{petersen2021pedagogical} \\

\textbf{\begin{tabular}[c]{@{}c@{}}Learning Outcomes \\ (DE10)\end{tabular}}  &          \\

\begin{tabular}[c]{@{}c@{}}Social Interaction \\ and Collaboration \end{tabular}           & \cite{sasikumar2024user}\cite{lee2024may}\cite{wong2023comparing}\cite{fidalgo2023magic}\cite{tian2023using}\cite{volonte2021effects}\cite{bozkir2021exploiting}\cite{aseeri2021influence}\cite{hsu2020design}\cite{wang2023designing}\cite{jin2024headsetoff}\cite{espositi2024room}\cite{mei2021cakevr}\cite{sykownik2022something}\cite{kimmel2023let}\cite{li2023we}\cite{wang2024socially}\cite{chen2024d}\cite{yakura2020enhancing}\cite{yoon2020evaluating}\cite{abramczuk2023meet}\cite{wei2023bridging}         \\

\begin{tabular}[c]{@{}c@{}}Spatial Understanding \\ and Navigation\end{tabular}                      & \cite{schott2024excuse}\cite{schott2023cardiogenesis4d}\cite{wijayanto2023comparing}\cite{bellgardt2023virtual}\cite{jetter2020vr}    \\

\begin{tabular}[c]{@{}c@{}}Presence, Embodiment \\ and Immersion\end{tabular}                        & \cite{yun2024exploring}\cite{bhargava2023empirically}\cite{schott2021vr}\cite{dickinson2021diegetic}\cite{hu2021abio}\cite{kim2022multi}\cite{thoravi2022dreamstream}\cite{thanyadit2023tutor}\cite{jicol2023realism}\cite{gomi2024inflatablebots}\cite{dufresne2024touching}\cite{gottsacker2021diegetic}\cite{schott2023vreal}\cite{wang2022realitylens}\cite{10.1145/3641825.3687753}\cite{10.1145/3686910}\cite{moss2024going}        \\

\begin{tabular}[c]{@{}c@{}}Emotional and \\ Psychological Aspects\end{tabular}            & \cite{unruh2023body}\cite{choudhary2023exploring}\cite{gu2022role}\cite{mal2023impact}\cite{voigt2021don}\cite{peck2021evidence}\cite{wolf2021embodiment}\cite{shigyo2024vr}\cite{bahng2020reflexive}\cite{jicol2021effects}\cite{shen2024legacysphere}\cite{luo2024exploring}\cite{10.1145/3492802}\         \\

\begin{tabular}[c]{@{}c@{}}Task Performance \\ and Efficiency\end{tabular}                           & \cite{reichherzer2022supporting}\cite{li2022synthesizing}\cite{wagner2021effect}\cite{li2020analysing}\cite{hamzeheinejad2021impact}\cite{michael2020race}\cite{lin2020architect}\cite{pohlmann2023you}\cite{kargut2024effects}\cite{bartl2022effects}\cite{10.1145/3641825.3687735}         \\

\begin{tabular}[c]{@{}c@{}}User Engagement, Motivation \\ and Learning\end{tabular} & \cite{bueno2021effects}\cite{wei2023feeling}\cite{wei2024hearing}\cite{koulouris2020me}\cite{petersen2021pedagogical}\cite{cao2023dreamvr}\cite{jin2023collaborative}      \\

\begin{tabular}[c]{@{}c@{}}Usability \\ and Interaction Design\end{tabular}            & \cite{yang2022hybridtrak}\cite{o2023re}        \\

\begin{tabular}[c]{@{}c@{}}Ethics \\ and Social Issues\end{tabular}             & \cite{tseng2022dark}         \\

\hline
\end{tabular}
}
\label{appendix-a}
\end{table}

\end{document}